\documentclass[lettersize,journal]{IEEEtran}
\usepackage{amsmath,amsfonts}
\usepackage{algorithmic}
\usepackage{algorithm}
\usepackage{array}
\usepackage[caption=false,font=normalsize,labelfont=sf,textfont=sf]{subfig}
\usepackage{textcomp}
\usepackage{stfloats}
\usepackage{url}
\usepackage{verbatim}
\usepackage{graphicx}
\usepackage{multirow}
\usepackage{multicol}
\usepackage{todonotes}
\usepackage{hyperref}
\usepackage{quantikz}

\begin{document}

\title{On the impact of key design aspects in simulated Hybrid Quantum Neural Networks for Earth Observation}

\author{Lorenzo~Papa,
        Alessandro~Sebastianelli,
        Gabriele~Meoni,
        and~Irene~Amerini,~\IEEEmembership{Member,~IEEE,}
\IEEEcompsocitemizethanks{
\IEEEcompsocthanksitem L. Papa and I. Amerini are with the  Department of Computer, Control and Management Engineering, Sapienza University of Rome, Italy, IT, 00185.
E-mail: [papa, amerini]@diag.uniroma1.it
\IEEEcompsocthanksitem A. Sebastianelli is with the $\phi$-lab at European Space Agency (ESA), Frascati, Italy, IT, 00078. 
E-mail: alessandro.sebastianelli@esa.int
\IEEEcompsocthanksitem G. Meoni is with the $\phi$-lab, ESA, Frascati, Italy, IT, 00078 and with the Advanced Concepts and Studies Office, ESA, Keplerlaan 1, 2201 AZ Noordwijk, Netherlands, NL
}
\thanks{The work has been developed during the visiting research period of L. Papa at $\phi$-lab European Space Agency (ESA), Frascati, Italy. 
}
\thanks{Manuscript received April 19, 2021; revised August 16, 2021.}
}

\markboth{Journal of \LaTeX\ Class Files,~Vol.~14, No.~8, August~2021}%
{Shell \MakeLowercase{\textit{et al.}}: A Sample Article Using IEEEtran.cls for IEEE Journals}

\IEEEpubid{0000--0000/00\$00.00~\copyright~2021 IEEE}

\maketitle

\begin{abstract}
Quantum computing has introduced novel perspectives for tackling and improving machine learning tasks. Moreover, the integration of quantum technologies together with well-known deep learning (DL) architectures has emerged as a potential research trend gaining attraction across various domains, such as Earth Observation (EO) and many other research fields. However, prior related works in EO literature have mainly focused on convolutional architectural advancements, leaving several essential topics unexplored.  Consequently, this research investigates through three cases of study fundamental aspects of hybrid quantum machine models for EO tasks aiming to provide a solid groundwork for future research studies towards more adequate simulations and looking at the post-NISQ era. More in detail, we firstly (1) investigate how different quantum libraries behave when training hybrid quantum models, assessing their computational efficiency and effectiveness. Secondly, (2) we analyze the stability/sensitivity to initialization values (i.e., seed values) in both traditional model and quantum-enhanced counterparts. Finally, (3) we explore the benefits of hybrid quantum attention-based models in EO applications, examining how integrating quantum circuits into ViTs can improve model performance. 
\end{abstract}


\section{Introduction}
\label{sec:introduction}
The advent of quantum computing has introduced revolutionary opportunities for tackling machine learning (ML) tasks from a new and powerful perspective. 
Quantum computing leverages the principles of superposition, entanglement, and quantum interference, which enable the processing of complex computations that are far beyond the capabilities of classical systems.
More in detail, in recent years, traditional ML approaches, especially deep learning (DL), have shown outstanding performances across several domains, including image recognition, natural language processing, and Earth Observation. 
However, as the researcher's pursuit of higher accuracy and efficiency continues to grow, the limitations of traditional computing become more evident.

As a result, the integration of quantum computing with traditional DL paradigms has emerged as a promising research frontier across various domains. 
This development is primarily driven by the ability of quantum algorithms to process and encode high-dimensional data in ways that classical systems find challenging, thereby offering the potential for improved accuracy and efficiency. 
Furthermore, quantum-enhanced models are capable of exploring larger solution spaces more effectively, which may lead to faster convergence training behaviors, improved generalization, and superior performance, especially in tasks involving large-scale and complex datasets.

\IEEEpubidadjcol
Consequently, following this research trend, several works on remote sensing data have been developed in order to investigate the use of such innovative, powerful technologies over EO tasks.
However, despite the growing interest in the combination of quantum computing (QC) and DL4EO, the majority of the existing research has been primarily focused on advancing hybrid models from an architectural perspective, i.e., focusing on (convolutional) encoding and/or quantum circuit components. 
Despite such significant advances, they represent just a portion of the broader challenges and potential associated with the application of quantum-enhanced models in EO tasks.
Consequently, building on previously related works and motivated by Zaidenberg et al. \cite{zaidenberg2021advantages}, this study intends to explore several key features that are central to advancing the field of hybrid quantum DL models and their applications in EO tasks. 
More in detail, the rationale of this work is threefold:
\begin{enumerate}
    \item Starting from Zaidenberg et al. \cite{zaidenberg2021advantages}, this study aims to evaluate the behavior of quantum computing libraries used to train quantum neural network (QNN) architectures.
    \item Investigate and compare the sensitivity performance of both quantized and non-quantized neural networks in response to different initialization (i.e., seed values). 
    Specifically, we will examine the convergence behavior of chosen architectures when subject to different starting conditions. 
    \item Explore the potential of hybrid quantum architectures by incorporating simple single quantum circuits into Vision Transformer (ViT) structures. 
    More in detail, the objective is to push the boundaries of the work proposed by Zaidenberg et al. \cite{zaidenberg2021advantages} assessing the performance of these novel HQViT models in Earth Observation (EO) tasks and comparing their behavior to their non-quantized counterpart. 
\end{enumerate}

However, while quantum computing has the potential to give new prospects when compared to traditional DL learning frameworks, it still faces several significant challenges and limitations.
Concerns include hardware stability, error rates, scalability, and the difficulty of developing effective quantum algorithms. 
Consequently, we have to consider that these challenges may limit the practical deployment of quantum-enhanced neural networks and have to be considered while assessing their potential in real-world applications.

Summarizing (1) by evaluating different quantum libraries, this research seeks to uncover potential performance discrepancies and challenges that may arise when implementing quantum-enhanced neural networks. 
The second point (2) is crucial for understanding how initialization impacts the stability and training efficacy of quantum and classical networks, as well as evaluating their sensitivity to initial parameter choices, which is a common concern in neural network training.
Finally, the third study (3) is motivated by the rising interest in hybrid quantum-classical techniques, which exploit quantum components to augment the capabilities of classical neural architectures in specific domains such as image classification and remote sensing.

As a result, while considering quantum limitations, this work intends to contribute to the growing body of knowledge on QNN by systematically investigating the interactions between quantum libraries, initialization values, and hybrid model structures, with a particular focus on their application to EO tasks.

The rest of this paper is organized as follows: Section \ref{sec:related_works} reviews the relevant literature on quantum and non-quantum deep learning approaches for Earth Observation (EO). Section \ref{sec:method} outlines the three case studies, highlighting their respective challenges. Section \ref{sec:implementation_details} provides a detailed description of the dataset and the implementation specifics required to replicate the reported experiments. Section \ref{sec:results_and_discussion} presents and analyzes the experimental results, while Section \ref{sec:conclusion} provides final thoughts and discusses future research directions.

\section{Related Works}
\label{sec:related_works}
The rapid advancements in both DL and quantum computing have generated significant interest in the EO domain in recent years. 
Consequently, we review recent related studies applied to EO tasks. 
This section provides a comprehensive overview of hybrid approaches, key challenges, and the potential of quantum computing in EO.

Zeng et al. \cite{zeng2020sub} (2020) laid the groundwork by introducing the Quantum Mechanism Effect Spectral Clustering (QMESC) model. 
Their model leverages quantum mechanics to tackle pixel mixture challenges in hyperspectral images, using Green’s function to accurately decompose mixed pixels and identify cluster centers with quantum potential energy. 

The following year, Zaidenberg et al. \cite{zaidenberg2021advantages} (2021) further advanced this field by developing a QNN model for remote sensing image classification using the EuroSAT dataset \cite{helber2019eurosat}. 
Their study emphasizes the speed and feasibility of QML for EO, showcasing performance on par with classical models. 
The author focuses on qubit decoherence and data processing on Noisy Intermediate-Scale Quantum (NISQ) devices, underscoring the need for improvements in data handling and model scalability for future applications.
In the same year, Otgonbaatar and Datcu \cite{otgonbaatar2021quantum} (2021) explored quantum annealing with a D-Wave quantum computer for feature selection in hyperspectral images. 
Their Mutual Information-based method identifies the most informative spectral bands, demonstrated on the Indian Pine dataset. 
Therefore, by employing quantum classifiers like Qboost, their approach achieved comparable or improved accuracy over classical methods, illustrating quantum annealing's potential for remote sensing data processing.
Sebastianelli et al. \cite{sebastianelli2021circuit} (2022) build upon \cite{zaidenberg2021advantages}, introducing a hybrid quantum convolutional neural network (HQCNN) that incorporates quantum layers within a classical CNN for enhanced land-use classification. 
Tested on the EuroSAT dataset, the authors show that HQCNN can improve traditional DL models by leveraging entanglement for improved classification accuracy. 
This work highlights the potential of quantum circuits for EO, paving the way for future applications with hybrid architectures.
Furthermore, Máté et al. \cite{mate2022beyond} (2022) proposed an ansatz-free optimization technique for quantum circuits, parameterizing circuits in the Lie algebra to simplify optimization and enhance training speed. 
This approach enables flexible exploration of quantum circuits, avoiding the constraints of fixed architectures. Tested on both toy and image classification tasks, their method demonstrates the computational advantages of unitary optimization, adding robustness to quantum machine learning models.
Expanding on hybrid quantum-classical approaches, Otgonbaatar et al. \cite{otgonbaatar2022quantum} (2022) investigated networks for large-scale EO data processing. 
They identified real-world problems suitable for quantum computing and proposed encoding strategies on NISQ devices. Their comparisons between hybrid models and conventional techniques underscore the potential for quantum computing to handle big data challenges, even amid hardware limitations.
Moreover, Gupta et al. \cite{gupta2022quantum} (2022) examined the integration of classical neural networks with Projected Quantum Kernel (PQK) features for Land Use and Land Cover tasks using Sentinel-2 data. 
They found that PQK significantly improved training accuracy, highlighting the advantages of QML in handling multispectral EO data. 
This study suggests promising avenues for future applications of quantum-enhanced features in remote sensing.

Further developments in 2023 saw Gupta et al. \cite{gupta2023potential} investigating PQK features for multispectral classification. They achieved substantial accuracy gains, underscoring the utility of quantum kernels for complex EO datasets. 
Chang et al. \cite{chang2023approximately} introduced Equivariant Quantum Convolutional Neural Networks (EquivQCNN), which leverage planar symmetries to enhance generalization and performance, particularly in data-limited scenarios, while highlighting the potential for symmetry-based quantum models in EO.
Furthermore, Nammouchi et al. \cite{nammouchi2023quantum} (2023) provided a comprehensive review of QML applications in climate change and sustainability, emphasizing quantum methods' potential in areas like energy systems and disaster prediction. They also discuss challenges with current quantum hardware, suggesting that QML could improve model accuracy and data processing efficiency in climate research, with potential expansions into modeling extreme events.
Moreover, Otgonbaatar et al. \cite{otgonbaatar2023quantum} (2023) firstly explored hybrid quantum transfer learning, combining classical VGG16 with QML for high-dimensional EO datasets. 
They compared real amplitude and strongly entangling quantum networks, finding that the latter often yielded better accuracy due to their local effective dimension, despite challenges related to limited quantum resources.
Subsequently, in another study, Otgonbaatar et al. \cite{otgonbaatar2023_1quantum} (2023) employed quantum-inspired tensor networks to enhance deep learning models for Earth science tasks. They focused on compressing physics-informed neural networks (PINNs) and improving the spectral resolution of hyperspectral images, achieving computational efficiency without compromising accuracy.

Recently, Fan et al. \cite{fan2024land} (2024) presented two HQCNNs for land cover classification using Sentinel-2 multispectral images. 
Their models combine quantum computing for feature extraction and classical methods for classification, achieving a performance boost over traditional CNNs. 
Similar to previous studies, also this research underlines the advantages of hybrid convolutional models in handling large EO datasets with improved accuracy and transferability.
Moreover, Meyer et al. \cite{meyer2024quantum} (2024) investigate a different approach by applying quantum reinforcement learning to cognitive synthetic aperture radar (SAR) data for ship detection in maritime monitoring. 
Their two-stage approach integrates variational quantum circuits for scene adaptation and resource optimization, demonstrating how quantum methods could enhance SAR systems' adaptability and efficiency in EO.

This timeline of advancements demonstrates the growing potential of quantum computing in EO, from quantum-enhanced clustering and feature selection to hybrid architectures and reinforcement learning. 
These studies collectively underscore the transformative impact quantum computing could have on EO, offering promising directions for future research and applications.
Consequently, this research study aims to build on previous knowledge exploring through three cases of study less-investigated quantum aspects, i.e., focusing on quantum libraries, sensitivity, and attention-based quantum structures.

\section{Cases of Study}
\label{sec:method}
This section presents the three key areas of investigation in this study: quantum libraries in Section \ref{subsec:quantum_libraries}, model robustness \ref{subsec:seed} in Section \ref{subsec:seed}, and architectural design in Section \ref{subsec:architectures}. 
More in detail, we first describe the quantum computing libraries utilized for training quantum neural networks, examining their strengths and limitations. 
Next, we look into the sensitivity (sensitivity to initialization) of both quantized and non-quantized models by analyzing the importance of the impact of different random seed initializations. 
Finally, we define the architectures employed in this study and introduce the novel hybrid quantum ViTs.

\subsection{Quantum Libraries}
\label{subsec:quantum_libraries}
Quantum computing has emerged as a revolutionary field capable of solving complex problems that are intractable for conventional computers, i.e., challenges that are too difficult or highly time-consuming.
This behavior is mainly motivated by the fact that differently, unlike classical computers, which process information using bits (0s and 1s), quantum computers use quantum bits (qubits), allowing them to perform several calculations simultaneously.  
As researchers and developers explore this new frontier, various quantum computing libraries have been developed to facilitate the design, simulation, and execution of quantum algorithms.
In this domain, two well-known frameworks are Qiskit and PennyLane, each offering specific features and capabilities that cover various elements of quantum computing and its integration with traditional machine learning techniques.
\noindent
\begin{center}
    - - - \\
\end{center}

\noindent
\textbf{Qiskit} has been developed by IBM company; it is a comprehensive framework that offers a wide range of tools for designing, simulating, and executing quantum circuits.
One of its key features is the ability to access real quantum hardware through the IBM Quantum platform, which allows for practical experimentation. 
Qiskit's modular architecture enables users to work with specific components, such as Qiskit Terra for circuit creation, Qiskit Aer for simulation, and Qiskit Ignis for error mitigation; this structure covers a wide range of applications/requirements.
Additionally, Qiskit also benefits from extensive documentation and a large community, which facilitates learning and troubleshooting. 
However, when compared with PennyLane, due to the needed interaction between multiple components,  Qiskit could be trickier, requiring a substantial time and effort investment. 
Furthermore, while Qiskit provides access to quantum devices, the performance and availability can be influenced by hardware limitations, such as qubit count and coherence time. 
Then, as quantum circuits scale in size, managing complexity and ensuring effective execution on available hardware becomes increasingly challenging.

\vspace{1em}
\noindent
\textbf{PennyLane} has been developed by Xanadu company; it is specifically designed for quantum/traditional computations, and it integrates with popular machine learning libraries like PyTorch and TensorFlow. 
This integration allows for the efficient development of hybrid quantum models, which is one of PennyLane's key advantages. 
The framework supports differentiable quantum programming, enabling users to optimize quantum circuits alongside traditional neural networks based on backpropagation techniques. 
Moreover, PennyLane's design is flexible, allowing for multiple quantum hardware platforms and simulators and providing researchers with a wide range of experimental choices. 
However, PennyLane also has its drawbacks. 
For instance, even if the framework supports several backends, users may find limited access to real quantum devices, depending on the platform they choose. 
Additionally, the learning curve associated with understanding the hybrid model concept and differentiable programming can pose challenges in the model's convergence behaviors.

\noindent
\begin{center}
    - - - \\
\end{center}

In summary, each quantum library has its advantages and disadvantages that pose challenges in its implementation and usage. 
Moreover, the development of hybrid models that effectively leverage both classical and quantum components could not come without limitations, even with the support of such powerful libraries. 
Furthermore, the research field of quantum computing is evolving rapidly, necessitating continuous learning and adaptation to new features and best practices within these libraries.
Motivated by previous claims, in this first case of study, we aim to investigate the practical usage and the model's convergence behavior in hybrid quantum settings in order to understand the framework's strengths and weaknesses.

\subsection{Sensitivity to initialization}
\label{subsec:seed}
In DL, the concept of sensitivity to initialization refers to how the initial weights and biases affect the training dynamics, convergence rate, and final performance of the model. 
Stability, on the other hand, measures the consistency of a model's performance across different runs under varying initial conditions, such as different random seeds. 
Therefore, the concept of sensitivity to initialization is crucial in DL research scenarios, as it influences how the optimization process navigates the high-dimensional loss landscape. 
More in detail, the weights of neural network architectures are typically initialized randomly or following a given distribution guided by a random value (seed), such as Normal, Uniform, and many others.
From a mathematical point of view, given a loss function \(\mathcal{L}(\theta)\), where \(\theta\) represents the parameters of the model, in a standard training procedure, such function is minimized (or maximized) through an optimization algorithm.
We following report (Equation \ref{eq:param_update}) how the parameters are updated at each time step ($t+1$):
\begin{equation}
    \theta_{t+1} = \theta_t - \eta \nabla \mathcal{L}(\theta_t)
    \label{eq:param_update}
\end{equation}

We indicate with $\eta$ the learning rate, and with $\nabla \mathcal{L}(\theta_t)$ the gradient of the loss function with respect to the parameters at time step $t$. 
Building on this formulation, the second case study of this work aims to explore $\theta_0$, i.e., the initialization of $\theta$ at time $t_0$. 
This focus is motivated by the fact that DL models may converge to suboptimal (local) minima or present divergent behavior due to inadequate initialization, particularly in complex loss surfaces characterized by local minima and saddle points.
Generally speaking, we investigate and compare classical DL models with their quantum-enhanced counterparts, detailed in the next section, under various initialization values/conditions. 
The objective is to examine the stability and convergence behaviors of novel techniques in comparison to traditional approaches within convolutional and transformer structures.
This concern is, in fact, particularly relevant in the context of hybrid quantum models, where the interplay between classical and quantum layers may present specific challenges in ensuring stable and reliable convergence. 
More in detail, in our scenario, the quantum layer/circuit is added to a conventional convolutional or transformer architecture in order to increase the feature space by leveraging quantum properties and potentially enhancing the model's performances.
However, such a layer may also introduce additional sensitivity and variability. 
These factors, together with quantum noise and gate fidelity, may significantly impact the stability of such hybrid models.
Mathematically speaking, the output state of the quantum layer, reported in Equation \ref{eq:out_quantum}, can be expressed as a unitary transformation applied to the input state vector $|\psi_{in}\rangle$:
\begin{equation}
    |\psi_{\text{out}}\rangle = U(\theta) |\psi_{\text{in}}\rangle
    \label{eq:out_quantum}
\end{equation}

where \(U(\theta)\) is a unitary operator parameterized by \(\theta\), representing the sequence of quantum gates applied to the input state.

Thus, the comparative analysis of classical and quantum-enhanced models in this study aims to provide insights into the benefits and trade-offs associated with quantum integration into traditional architectures. 
Moreover, by examining convergence behaviors across multiple seed values, the study aims to explore robustness and stability while offering guidance for the development of future quantum-classical hybrid neural networks.


\subsection{Architectures}
\label{subsec:architectures}
\begin{figure}[t]
    \centering
    \includegraphics[width=\linewidth]{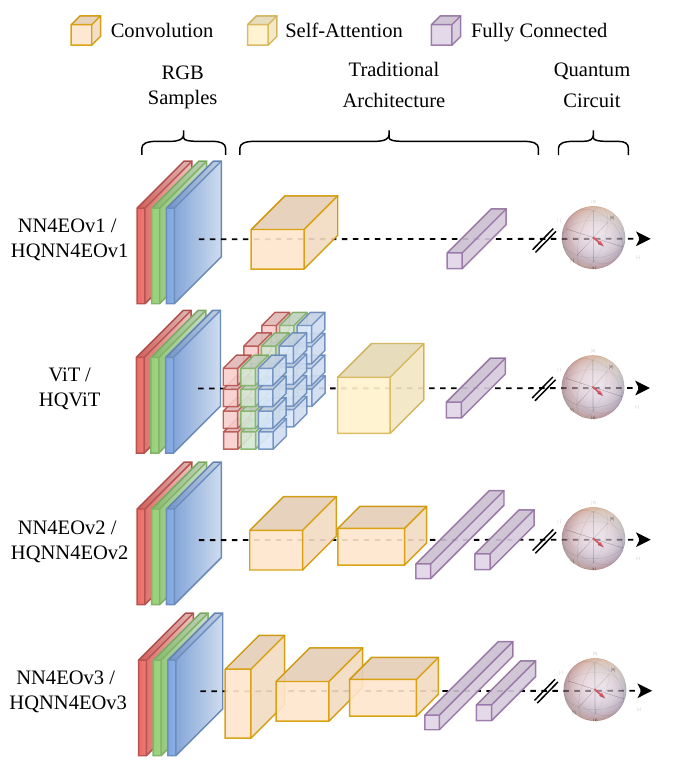}
    \caption{Graphical representation of the three reference models employed in this research study. 
    Each traditional architecture, i.e., NN4EOv1, NN4EOv2, NN4EOv3, and ViT, is composed of a sequence of convolutional/self-attention layers (in orange/yellow) in addition to fully connected layers for classification. 
    Differently, the quantum models, i.e., HQNN4EOv1, HQNN4EOv2, HQNN4EOv3, and HQViT, are developed by stacking a quantum circuit to the fully connected layers of traditional models.
    Within the same architectural design, double lines are used to distinguish between traditional and hybrid designs, while a Bloch sphere represents the single qubit circuit.
    }
    \label{fig:compared_models}
\end{figure}

In this last section, we formally describe quantized and traditional architectures while introducing innovative hybrid quantum Vision Transformers (ViTs), which, to the best of our knowledge, are being employed for the first time in EO tasks. 
Specifically, this study examines four couple of architectural structures: three convolutional architectures, namely NN4EOv1, NN4EOv2, NN4EOv3 in their traditional forms, and their quantized counterparts, HQNN4EOv1, HQNN4EOv2, and HQNN4EOv3 which has been originally extracted from Zaidenberg et al. \cite{zaidenberg2021advantages} and reduced in terms of number of convolutional operation to understand their behavior
Differently, the latter ViT-based structure is referred to as ViT and HQViT. 
These architectural structures are graphically represented in Figure \ref{fig:compared_models} and following described.

Before going into the details of each architectural structure, we formally introduce their fundamental elements, i.e., the convolution operation employed in CNN and NN4EO, the self-attention mechanism used in ViT, and the elementary quantum layer utilized in their hybrid quantum configurations.

\noindent
\begin{center}
    - - - \\
\end{center}

\noindent
The \textbf{2D Convolution} operation is the foundational operation in image processing and a key component of well-established CNN architectures. This operation involves a filter (kernel), which slides over the input image to produce an output feature map. The primary objective of convolution is to extract an image's features, such as edges, textures, or patterns.
More in detail, given an input image \( I \) and a kernel \( K \), the convolution produces an output feature map \( O \). 
Thus, the value of the output \( O \) at the pixel position \( (i, j) \) is computed as follows:
\begin{equation}
    O(i, j) = \sum_{m=-a}^{a} \sum_{n=-b}^{b} I(i + m, j + n) \cdot K(m + a, n + b)
\end{equation}

Where \( I(i, j) \) represents the pixel value at the coordinates \( (i, j) \) in the input image, while \( K(m, n) \) denotes the value at position \( (m, n) \) within the kernel, which has dimensions of \( (2a + 1) \times (2b + 1) \). 
The parameters \( a \) and \( b \) represent the half-widths of the kernel in the vertical and horizontal directions, respectively. 
Consequently, as the kernel slides along the image, it computes a weighted sum of the pixel values covered by the kernel. 
This process effectively captures local patterns and translates the original image into a more abstract representation.
Such a procedure enables subsequent convolutional layers to learn and extract increasingly complex features. 
To summarize, the kernel and its parameters significantly influence the performances of the convolution operation, as well as the types of features/information extracted from the image.

\vspace{1em}
\noindent
The \textbf{self-attention mechanism}, introduced by Vaswani et al. \cite{vaswani2017attention}, is the key component of the attention block employed in ViT architectures. 
This mechanism is specifically designed to capture long-range relationships in image data by operating on embedded images or feature patches. 
In particular, the self-attention operation allows each patch to relate to all others within the sequence, thereby increasing the DL model's receptive field with respect to conventional local convolutional operations.
Mathematically, given an input sequence of embedded patches, self-attention computes three matrices: the query ($Q$), key ($K$), and value ($V$).
Subsequently, as detailed in Equation \ref{eq:self-attention}, the self-attention is computed performing the dot-product interactions between queries and keys, scaled by the dimensionality $\sqrt{d_k}$, followed by a softmax operation in order to generate attention scores, which are then applied to the values ($V$).
\begin{equation}
    A(Q, K, V) = \text{Softmax} \left(\frac{Q \cdot K^T}{\sqrt{d_k}}\right) \cdot V
    \label{eq:self-attention}
\end{equation}

Such a formula computes the dot-product interactions between queries and keys, scaled by the dimensionality $\sqrt{d_k}$, followed by a softmax operation to generate attention scores, which are then applied to the values. 
However, as detailed in Papa et al. \cite{papa2024survey}, the time and memory complexity of this operation is $\mathcal{O}(n^2)$ due to the quadratic cost of computing $A(Q, K, V)$ making this operation particularly powerful but computationally expensive for large input sizes.

Furthermore, this elementary operation can be parallelized into a multi-head self-attention (MSA) mechanism, in which multiple self-attention layers are executed simultaneously. 
This solution allows the model to focus on different areas/characteristics of the input features at the same time. 
More in detail, given the input features \(X\), the output features \(X_{out}\) resulting from the execution of an attention block can be mathematically formulated as follows:
\begin{equation}
\begin{split}
    & X_{MSA} = \text{Norm}(\text{MSA}(X, X)) + X \\
    & X_{out} = \text{Norm}(\text{FNN}(X_{MSA})) + X_{MSA}
\end{split}
\end{equation}

Here, Norm denotes a normalization process, whereas FNN indicates a feed-forward network.

\vspace{1em}
\noindent
The \textbf{quantum layer} is a key component in quantum neural networks.
It consists of a sequence of quantum gates that perform unitary transformations on qubits, allowing the manipulation and entanglement of quantum states. 
Quantum layers can also be designed to operate as the quantum equivalent of classical neural network layers.
However, similar to binary architectures, such a qubit-based layer enables the encoding, processing, and transformation of input data within quantum circuits.
Here, the fundamental elements of the elementary quantum circuit used in this work are described. 
Specifically, the structure of the quantum circuit, as following illustrated, has been derived from Zaidenberg et al. \cite{zaidenberg2021advantages}.
\begin{center}
    \begin{quantikz}
       \vert 0 \rangle \hspace{2mm} & \gate{H} & \qw & \gate{R_y(\theta)} & \qw & \meter{}
    \end{quantikz}
\end{center}

Generally speaking, the core of quantum computing is the concept of qubit, a two-level quantum system that can be represented on the Bloch sphere. 
The Bloch sphere provides a geometric representation of the qubit's state, where any point on the sphere corresponds to a valid qubit state. 
The north pole represents the state ($\vert 0\rangle$), and the south pole represents ($\vert 1\rangle$); mathematically, a qubit can be expressed as a linear combination of its basis states as reported in Equation \ref{eq:qubit}.
\begin{equation}
    \vert \psi\rangle = \alpha \vert 0 \rangle + \beta \vert 1 \rangle
    \label{eq:qubit}
\end{equation}

Here, ($\alpha$) and ($\beta$) are complex coefficients satisfying the normalization condition ($|\alpha|^2 + |\beta|^2 = 1$).
Moreover, following the single-qubit circuit previously reported, several key operations are performed: (1) the qubit is initialized to a specific state, typically \(|0\rangle\).
Then, (2) the Hadamard ($H$) gate is used in order to create a superposition state. 
The action of the Hadamard gate on the basis states is defined as:
\[
H|0\rangle = \frac{1}{\sqrt{2}}(|0\rangle + |1\rangle), \quad H|1\rangle = \frac{1}{\sqrt{2}}(|0\rangle - |1\rangle)
\]
Where the matrix representation of the Hadamard gate is:
\[
H = \frac{1}{\sqrt{2}} \begin{pmatrix}
1 & 1 \\
1 & -1
\end{pmatrix}
\]

Subsequently, (3) a rotation gate ($R_y(\theta)$) allows manipulation of the qubit's state. 
In our case scenario, the rotation around the Y-axis of the Bloch sphere is given by the following formula:
\[
R_y(\theta) = e^{-i \frac{\theta}{2} Y} = \cos\left(\frac{\theta}{2}\right) I - i \sin\left(\frac{\theta}{2}\right) Y
\]
where \(Y\) is the Pauli-Y matrix:
\[
Y = \begin{pmatrix}
0 & -i \\
i & 0
\end{pmatrix}
\]

Finally, (4) the qubit's state is measured by collapsing its superposition into one of the basis states, i.e., the probability of measuring state \(|0\rangle\) or \(|1\rangle\) is given by:
\[
P(0) = |\alpha|^2, \quad P(1) = |\beta|^2
\]
\noindent
\begin{center}
    - - - \\
\end{center}

\noindent
Once the fundamental elements of each compared architecture have been introduced, we discuss and present the four architectural structures along with their respective quantum configurations.
More in detail, we leverage three convolutional and a ViT model.
A block diagram representation of these networks and their quantum counterparts are reported in Figure \ref{fig:compared_models}.
As can be noticed, all the architecture leverages fully connected layers in order to perform the final classification.
More in detail, the three convolutional architectures are composed of concatenations of convolutional blocks.
Each block is composed of a two-dimensional convolution with $5\times5$ kernel, followed by a $2\times2$ max pooling layer and a ReLU activation function.
Additionally, following Zaidenber et al. \cite{zaidenberg2021advantages}, in NN4EOv2 and NN4EOv3, two fully connected layers in which the first one is used to match the output flattened features from the previous encoding part and compact the information into $64$ output neurons, while, the second layer outputs the binary classification probability through a single neuron.
Differently, in NN4EO, and similarly to the ViT-based model, a single fully connected layer is used.
However, the CNN-based models differ for the number of subsequent convolutional blocks as illustrated in Figure \ref{fig:compared_models} (orange blocks), i.e., NN4EOv1, NN4EOv2, and NN4EOv3 are respectively composed by one, two and three convolutional blocks counting respectively $6.6K$, $18K$ and $68K$ trainable parameters.
Furthermore, the ViT model implemented for this study has been intentionally designed in order to maintain a simple architecture because the main objective of the third study proposed in this work is not to develop a highly complex model but rather to demonstrate the potential effectiveness of integrating quantum circuits with the ViT structure for EO tasks. 
More in detail, the input image is divided into $8 \times 8$ patches, which are then processed through a Multi-Head Self-Attention (MSA) layer with two attention heads, and finally fed into a single fully connected layer that takes the encoded features as input and return a single prediction with a single neuron.
This minimalistic design ensures a lightweight model architecture resulting in less than $34K$ trainable parameters.

On the other hand, quantum models, i.e., HQNN4EOv1, HQNN4EOv2, HQNN4EOv3, and HQViT, leverage the same architectural structure as their traditional counterparts, i.e., NN4EOv1, NN4EOv2, NN4EOv3, and ViT respectively, while adding the previously introduced single-qubit circuit for the final classification stage. 
The objective of this integration is to introduce quantum processing capabilities into traditional models, aiming to exploit quantum effects such as superposition and entanglement to potentially enhance classification performance on EO tasks.

\section{Experimental Setup}
\label{sec:experimsental_setup}
In this section, we detail the experimental setup used to evaluate the studies that have just been described in the EO domain.
The experimental setup is divided into two parts.
Firstly, we outline in Section \ref{sec:data}, the characteristics and prepossessing steps applied to the training dataset. 
Then, in Section \ref{sec:implementation_details}, we describe the implementation details, including the software libraries, training protocols, and hyperparameters used to train and evaluate the models. 

\subsection{Training Dataset}
\label{sec:data}
The study was conducted using the EO application scenario. 
More in detail, quantum architectures have been investigated in order to tackle the image classification task, specifically the identification of scenes in the EuroSat dataset \cite{helber2019eurosat}. 
This dataset is composed of Sentinel-2 data covering 13 spectral bands and is divided into $10$ classes, with a total of $27000$ labeled and georeferenced images. 
Moreover, following the training protocol proposed in Zaidenberg et al. \cite{zaidenberg2021advantages}, and in order to simplify the task given the innovative use of hybrid quantum vision transformers in the research field of EO, the number of classes has been reduced to two, resulting in multiple binary classification tasks. 
Precisely, at training time, the dataset has been subsequently split into training and validation sets with a division factor of $20\%$. 
Out of the 13 available bands, only the RGB bands have been selected.

\subsection{Implementation Details}
\label{sec:implementation_details}
The study has been implemented using PyTorch\footnote{The source code, corresponding pre-trained weights, and Docker images are available at the following GitHub repository: \href{https://github.com/...}{https://github.com/...}} (v12.4.1) deep learning API.
All models have been trained from scratch, following the training protocol outlined in \cite{zaidenberg2021advantages}, while the Binary Cross Entropy loss function has been used in order to perform the binary classification across all possible pairwise combinations of the 10 dataset's classes.
We identify such classes with numbers ranging from $0$ to $9$, which correspond to highway (0), forest (1), sea lake (2), herbaceous vegetation (3), river (4), industrial (5), residential (6), pasture (7), permanent crop (8), and annual crop (9).
Specifically, Adam optimizer \cite{kingma2014adam} has been employed with $\beta_1 = 0.9$, $\beta_2 = 0.999$, and an initial learning rate of $0.0001$ for a total of $20$ epochs with a batch size of $1$, and no data augmentation techniques applied to the training dataset.
Additionally, preliminary studies involving quantum circuits have been conducted using Qiskit (v1.2.0), while Pennylane (v0.28.0) has been employed to facilitate GPU support for quantum computations.
For robustness investigations, multi-start experiments has been performed using $k=10$ distinct seed values, specifically: $0$, $12$, $123$, $1000$, $1234$, $10000$, $12345$, $100000$, $123456$, $1234567$. 
Once the training phase has been concluded, we quantitatively evaluate the trained models using the accuracy metric ($Acc$), which is widely adopted in the literature.
Moreover, we evaluate the stability of reference models through the accuracy variance ($\sigma^2(Acc)$) across the $k$ training/seeds, as reported below.
\[
\sigma^2(Acc) = \frac{1}{k} \sum_{i=1}^{k} (Acc_i - \overline{Acc})^2
\]
Where $Acc_i$ is the accuracy performance of the $i$-th run, and $\overline{Acc}$ is the mean accuracy across all runs. 
For completeness, we remind that the lower the variance, the higher the stability, i.e., a highly stable model will exhibit minimal accuracy variability and lower variance, suggesting that the model's training dynamics are robust to random factors.

\section{Results and Discussion}
\label{sec:results_and_discussion}
This section will quantitatively analyze and compare the performance of eight models, namely four quantum and their respective non-quantum configurations. 
More in detail, Section \ref{subsec:quantum_lib} compares well-known quantum libraries to investigate their impact on QNNs training. 
Subsequently, in Section \ref{subsec:seeds}, the impact of varying initialization values on model performances and stability will be analyzed. 
Lastly, Section \ref{subsec:vit_qvit} will present a comparative analysis between HQViT and ViT for EO classification tasks.

\subsection{Comparison of Quantum Libraries}
\label{subsec:quantum_lib}
In this first set of experiments, we are going to compare the different performances of well-known quantum libraries. 
As introduced in Section \ref{sec:implementation_details}, we train four reference hybrid quantum models using Qiskit (v1.2.0) and PennyLane (v0.28.0) versions based on the same training configuration and a fixed seed value equal to 1699806.
Due to the high number of experiments, we report the obtained results in the attached Appendix.
More in details we report Tables \ref{tab:hqnn4eov1_quiskit}, \ref{tab:qcnn_Qiskit}, \ref{tab:qnn4eo_Qiskit}, and \ref{tab:qvit_Qiskit} respectively for HQNN4EOv1, HQNN4EOv2, HQNN4EOv3, and HQViT Qiskit configuration and in Tables \ref{tab:hqnn4eov1_pennylane}, \ref{tab:qcnn_pennylane}, \ref{tab:qnn4eo_pennylane}, and \ref{tab:qvit_pennylane} for the same architectures in the PennyLane configuration.
Moreover, in order to give a broader overview, we report in Table \ref{tab:Qiskit_vs_pennylane} a summary of the performed tests presenting a comparison between quantum models trained using the two previously introduced quantum computing libraries, i.e., Qiskit and PennyLane.
More in detail, we report the average accuracy ($\overline{Acc}$) and the average value in which the best model has been saved at training time  (\textit{k*}). 
The latter information can give us an overview of the amount of epochs needed for a model in order to reach convergence (a local minimum).
\begin{table}[h]
    \centering
    \caption{Quantum models comparison over Qiskit and PennyLane libraries}
    \begin{tabular}{ c||c|c||c|c }
        \multirow{2}*{Models} & \multicolumn{2}{c||}{Qiskit} & \multicolumn{2}{c}{PennyLane} \\ 
         & $\overline{Acc}$ & \textit{k*} & $\overline{Acc}$ & \textit{k*} \\ \hline
        HQNN4EOv1 & \textbf{91.93} & 16.31 & 91.80 & \textbf{15.53} \\ \hline
        HQNN4EOv2 & 92.35 & 16.36 & \textbf{92.51} & \textbf{16.11} \\ \hline
        HQNN4EOv3 & \textbf{93.45} & 15.89 & 93.15 & \textbf{15.46} \\ \hline
        HQViT & \textbf{87.95} & 16.25 & 87.77 & \textbf{16.20} \\ \hline
    \end{tabular}
    \label{tab:Qiskit_vs_pennylane}
\end{table}

Based on the reported results, both Qiskit and PennyLane libraries exhibit strong performance across all models, with only minor variations in accuracy and \textit{k*}. 
For instance, in the HQNN4EOv2 model, PennyLane achieves slightly better results in both accuracy, equal to $92.51\%$ and computational efficiency $\textit{k*} = 16.11$ compared to Qiskit, which achieves an accuracy of $92.35\%$ and a slightly higher \textit{k*} value equal to $16.36$. 
Differently, HQNN4EOv1, HQNN4EOv3, and HQViT models present a different scenario, where Qiskit slightly surpasses PennyLane in terms of accuracy, achieving the highest score. 
However, also in this scenario, PennyLane remains competitive with a close accuracy of $91.80\%$, $93.15\%$, and $87.77\%$ with respect to $91.93\%$, $93.45\%$, and $87.77\%$ respectively for HQNN4EOv1, HQNN4EOv3, and HQViT while achieving better convergence performances with a lower \textit{k*} epochs needed for Qiskit full convergence. 
These results suggest that while Qiskit performs slightly better in terms of accuracy in certain cases, PennyLane consistently demonstrates faster convergence behavior.

Furthermore, from a more detailed analysis of the tables presented in the Appendix, i.e., when comparing each pair of trained classes among the four-compared quantum models, the obtained results reveal that out of the 184 training sessions, i.e., 46 possible binary class configurations for each quantum-enhanced model, Qiskit and PennyLane obtains similar performances.
More in detail, across the 184 training sessions, Qiskit and PennyLane achieve the same performances, i.e., Quiskit outperforms PennyLane in $45.6\%$ ($84/184$) instances, PennyLane outperforms Quiskit in $44.6\%$ ($84/184$), while in $16$ sessions, both frameworks yield identical accuracy results.

In conclusion, both Qiskit and PennyLane perform well in terms of accuracy for EO classification tasks. 
However, PennyLane shows a potential advantage in computational efficiency, making it a valuable tool for scaling quantum models in resource-constrained environments.
The latter assumption is motivated by the fact that PennyLane achieves a constant advantage in terms of \textit{k*}, highlighting its potential for more efficient execution, especially when dealing with larger quantum circuits or more complex tasks. 
Additionally, PennyLane’s integration with PyTorch and its support for GPU acceleration further enhance its suitability for hybrid quantum-classical learning. These features suggest that PennyLane may be more advantageous in contexts where computational resources are limited, or efficiency is a key priority.

\subsection{Study on the Stability Towards Initialization Values}
\label{subsec:seeds}
In this section, which is related to the second case study of this work, we investigate and compare the stability and estimation performances of reference models.
Similar to the previous section, due to the extensive number of conducted experiments, we report the average class-wise results in the Appendix.
Specifically, the results for NN4EOv1, NN4EOv2, NN4EOv3, and ViT are detailed in Tables \ref{tab:variance_accuracy_nn4eov1}, \ref{tab:variance_accuracy_cnn}, \ref{tab:variance_accuracy_nn4eo}, \ref{tab:variance_accuracy_vit} respectively. 
Similarly, the outcomes for HQNN4EOv1, HQNN4EOv2, HQNN4EOv3, and HQViT are provided in Tables \ref{tab:variance_accuracy_hqnn4eov1}, \ref{tab:variance_accuracy_qcnn}, \ref{tab:variance_accuracy_qnn4eo}, and \ref{tab:variance_accuracy_qvit}, respectively. 
However, in order to provide a more general overview, we show in Table \ref{tab:Stability_and_MAccuracy} a summary of all the experiments, reporting the mean accuracy ($\overline{Acc}$) and mean-variance ($\overline{\sigma}^2$) across all classes over the $k=10$ seeds.

\begin{table}[h]
    \centering
    \caption{Model's comparison over $k$ initialization values}
    \begin{tabular}{ c||c|c||c|c }
        \multirow{2}*{Models} & \multicolumn{2}{c||}{traditional} & \multicolumn{2}{c}{quantum} \\ 
         & $\overline{Acc}$ & $\overline{\sigma}^2$ & $\overline{Acc}$ & $\overline{\sigma}^2$ \\ \hline
        NN4EOv1 / & \multirow{2}*{90.87} & \multirow{2}*{3.85} & \multirow{2}*{\textbf{90.90}} & \multirow{2}*{\textbf{3.25}} \\
        HQNN4EOv1 & & & & \\ \hline
        NN4EOv2 / & \multirow{2}*{\textbf{92.66}} & \multirow{2}*{\textbf{4.25}} & \multirow{2}*{92.56} & \multirow{2}*{4.44} \\
        HQNN4EOv2 & & & & \\ \hline
        NN4EOv3 /  & \multirow{2}*{93.00} & \multirow{2}*{2.72} & \multirow{2}*{\textbf{93.47}} & \multirow{2}*{\textbf{2.45}} \\ 
        HQNN4EOv3 & & & & \\ \hline
        ViT /  & \multirow{2}*{88.37} & \multirow{2}*{\textbf{3.47}} & \multirow{2}*{\textbf{88.78}} & \multirow{2}*{7.77} \\
        HQViT & & & & \\ \hline
    \end{tabular}
    \label{tab:Stability_and_MAccuracy}
\end{table}

Based on the obtained results, it can be noted that quantum-based models, specifically HQNN4EOv3 and HQViT, demonstrate advantages in terms of accuracy, achieving mean accuracy of $93.47\%$ and $88.78\%$, respectively, i.e., a $0.5\%$ boost when compared to their traditional counterparts.
Moreover, HQNN4EOv1 is also able to achieve small improvements with respect to its traditional variance.
This improvement may indicate that hybrid quantum models, even if with a small boost, can enhance model performance.
However, even if the improvement is limited, the quantum model is able to obtain higher estimations with a lower variance compared with its traditional configuration.
Similarly, the HQNN4EOv3 model not only shows superior accuracy but also exhibits reduced variance compared to its traditional version.
Such results may suggest that quantum-enhanced models can provide more consistent and stable performance across multiple initialization values.
However, it is important to acknowledge that the benefits of quantum models are not uniform across all compared architectures. 
For instance, HQViT, while showing improved prediction performances, achieves a higher variance when compared with its traditional counterpart.
This observation may underscore the need for careful parameter tuning when incorporating quantum elements. 
However, despite these challenges, the results reported in Table \ref{tab:Stability_and_MAccuracy}, shows that even the simple integration of a single qubit can yield to performance gains; suggesting that quantum layers, even in their simplest forms, can enhance traditional models.

In summary, we can assess that a careful design of the initialization and optimization strategies is essential to mitigate instability and achieve reliable performance.

\subsection{Towards Hybrid Quantum Vision Transformers for Earth Observation}
\label{subsec:vit_qvit}
In this section, we report the quantitative evaluation of experiments performed for the third case study of this work.
More in detail, we investigate the potential of HQViT architectures for EO tasks by comparing the performance estimation of a traditional ViT model with its quantum-enhanced counterpart; both the models have been detailed in Section \ref{sec:method}. 
The objective of this study, inspired by Zaidenberg et al. \cite{zaidenberg2021advantages} on CNN models, is to determine whether the integration of quantum circuits, even in their simplest structure, can positively contribute over traditional ViT approaches.

However, do the the large amount of performed experiments, we report the class-wide results in the Appendix in Tables \ref{tab:accuracy_matrix_vit} and \ref{tab:accuracy_matrix_qvit}.
Moreover, in order to give a faster look at the model's performances, we can refer to Tabe \ref{tab:Stability_and_MAccuracy}.

Based on the obtained results, it can be noticed that the average accuracy ($\overline{Acc}$) of the HQViT model is marginally higher ($88.78$) when compared to the traditional ViT (88.37) model.
This finding may indicate a present, albeit modest, improvement in the performance of the quantum-enhanced ViT structure. 
Consequently, the results suggest that even a minimal quantum integration can introduce qualitative improvements, potentially paving the way for more sophisticated and efficient quantum-augmented models. 

In conclusion, this third research study and respective set of experiments computed over minimal ViT-based architectural setups is thought of as proof-of-concept in order to highlight the potential of quantum computing in machine learning models. 
Moreover, the HQViT model shows that quantum-enhanced vision transformers can positively influence ViT-based models, encouraging future research into advanced quantum architectures and their integration into deep learning frameworks.

\section{Conclusions and Future Worls}
\label{sec:conclusion}
This study investigates less-explored aspects of quantum DL applications for EO tasks.
More in detail, building upon Zaidenberg et al. \cite{zaidenberg2021advantages}, three cases of study are investigated.
Firstly, we compare the convergence behavior of well-known quantum libraries, i.e., Quiskit and PennyLane, in order to understand their potential in training hybrid quantum models.
This first case of the study reveals that both libraries provide benefits for QNN models achieving comparable classification performances and convergence behaviors; however, PennyLane easily integrates with PyTorch GPU libraries, which is advantageous for researchers.
Secondly, we investigate the sensitivity/stability of quantum and traditional counterparts with respect to the initialization values (seeds).
This second case of the study reveals that both types of architecture need a careful design of the initialization hyperparameters in order to mitigate possible instabilities; however, over $k=10$ different trials, quantum models show higher (averaged) accuracy values with comparable (averaged) variance. 
These results underline the effective contribution of quantum structures into hybrid architectures, even with elementary circuits, i.e., in our case, a single-bit module.
Finally, the third case study investigates the use of such a single qubit circuit embedded into a transformer-based architecture.
More in detail, by combining a simple (2 heads) multi-head attention layer with the previously introduced circuit, we show that, even with a higher variance due to the initialization values, the HQViT model is able to achieve an average boost of almost $0.5\%$ when compared with its traditional counterpart.
This finding pushes the boundaries of prior research on classical convolution-based models by demonstrating the advantages of hybrid quantum architectures in complex real-world applications like EO.

In summary, this study provides evidence that quantum computing libraries and quantum circuits may offer significant advantages even with simple DL architectural structures.
Additionally, the successful integration of quantum circuits into ViT models for EO tasks may open new research trends for further exploration. 
Consequently, building on such findings, future research may focus on investigating if more extensive quantum circuits may reduce the variance with respect to the initialization values while leading to more stable models and optimizing hybrid quantum ViT architectures with more architectural-oriented structures for EO tasks and more complex EO applications in order to take advantage of such kind of global processing with respect to convolutional-based models.

\bibliography{main}




\begin{table*}[h]
    \centering
    \footnotesize
    \caption{HQNN4EOv1 - Quiskit - Avg Test Accuracy 91.93 and best model saved at epoch 16.31}
\begin{tabular}{|c|c|c|c|c|c|c|c|c|c|c|}
    \hline
         & 0             & 1             & 2             & 3             & 4             & 5             & 6             & 7             & 8             & 9             \\ \hline
    0    & -             & 94.27 (17)    & 94.55 (17)    & 76.82 (20)    & 75.90 (20)    & 94.50 (19)    & 87.64 (17)    & 89.11 (16)    & 77.80 (20)    & 87.91 (18)    \\ \hline
    1    & 94.27 (17)    & -             & 89.33 (19)    & 97.33 (14)    & 88.27 (20)    & 99.91 (8)     & 99.17 (17)    & 92.00 (13)    & 97.64 (6)     & 96.67 (15)    \\ \hline
    2    & 94.55 (17)    & 89.33 (19)    & -             & 94.00 (20)    & 93.36 (19)    & 99.64 (19)    & 99.33 (7)     & 95.90 (18)    & 98.45 (18)    & 93.33 (16)    \\ \hline
    3    & 76.82 (20)    & 97.33 (14)    & 94.00 (20)    & -             & 88.00 (20)    & 95.00 (12)    & 90.67 (18)    & 92.80 (10)    & 70.00 (19)    & 87.17 (17)    \\ \hline
    4    & 75.90 (20)    & 88.27 (20)    & 93.36 (19)    & 88.00 (20)    & -             & 98.80 (19)    & 94.82 (16)    & 87.22 (20)    & 90.90 (18)    & 89.91 (15)    \\ \hline
    5    & 94.50 (19)    & 99.91 (8)     & 99.64 (19)    & 95.00 (12)    & 98.80 (19)    & -             & 86.55 (19)    & 99.67 (16)    & 96.90 (19)    & 99.55 (16)    \\ \hline
    6    & 87.64 (17)    & 99.17 (17)    & 99.33 (7)     & 90.67 (18)    & 94.82 (16)    & 86.55 (19)    & -             & 97.60 (16)    & 95.36 (15)    & 98.17 (14)    \\
\hline
    \end{tabular}
    \label{tab:hqnn4eov1_quiskit}
\end{table*}

\begin{table*}[h]
    \centering
    \footnotesize
    \caption{HQNN4EOv1 - PennyLane - Avg Test Accuracy 91.80 and best model saved at epoch 15.53}
    \begin{tabular}{|c|c|c|c|c|c|c|c|c|c|c|}
    \hline
         & 0             & 1             & 2             & 3             & 4             & 5             & 6             & 7             & 8             & 9             \\ \hline
    0    & -             & 94.00 (17)    & 94.18 (17)    & 77.00 (16)    & 74.50 (18)    & 94.30 (17)    & 87.18 (20)    & 89.22 (16)    & 77.90 (19)    & 87.64 (16)    \\ \hline
    1    & 94.00 (17)    & -             & 89.83 (20)    & 97.08 (20)    & 88.73 (19)    & 99.73 (4)     & 99.08 (17)    & 91.90 (11)    & 97.91 (2)     & 96.92 (15)    \\ \hline
    2    & 94.18 (17)    & 89.83 (20)    & -             & 93.75 (20)    & 92.45 (19)    & 99.64 (19)    & 99.17 (7)     & 95.60 (20)    & 98.18 (18)    & 93.67 (18)    \\ \hline
    3    & 77.00 (16)    & 97.08 (20)    & 93.75 (20)    & -             & 88.73 (20)    & 94.82 (19)    & 91.25 (16)    & 93.10 (20)    & 70.09 (19)    & 88.00 (15)    \\ \hline
    4    & 74.50 (18)    & 88.73 (19)    & 92.45 (19)    & 88.73 (20)    & -             & 98.20 (19)    & 93.91 (13)    & 87.00 (13)    & 90.40 (20)    & 89.55 (19)    \\ \hline
    5    & 94.30 (17)    & 99.73 (4)     & 99.64 (19)    & 94.82 (19)    & 98.20 (19)    & -             & 85.45 (19)    & 99.78 (18)    & 96.40 (20)    & 99.09 (17)    \\ \hline
    6    & 87.18 (20)    & 99.08 (17)    & 99.17 (7)     & 91.25 (16)    & 93.91 (13)    & 85.45 (19)    & -             & 97.50 (16)    & 95.36 (19)    & 98.83 (20)    \\
\hline
    \end{tabular}
    \label{tab:hqnn4eov1_pennylane}
\end{table*}

\begin{table*}[h]
    \centering
    \footnotesize
    \caption{HQNN4EOv2 - Qiskit - Avg Test Accuracy 92.35 and best model saved at epoch 16.38}
    \begin{tabular}{|c|c|c|c|c|c|c|c|c|c|c|}
    \hline
         & 0           & 1             & 2             & 3             & 4             & 5             & 6             & 7             & 8             & 9            \\ \hline
    0    & -           & 99.45 (20)    & 99.27 (20)    & 80.73 (17)    & 72.20 (18)    & 95.20 (16)    & 92.36 (20)    & 87.89 (16)    & 79.80 (20)    & 80.45 (18)   \\ \hline
    1    & 99.45 (20)  & -             & 87.42 (17)    & 97.83 (20)    & 96.18 (18)    & 100.00 (3)    & 100.00 (6)    & 92.40 (14)    & 99.45 (17)    & 97.92 (18)   \\ \hline
    2    & 99.27 (20)  & 87.42 (17)    & -             & 94.42 (19)    & 96.09 (19)    & 100.00 (9)    & 99.92 (17)    & 92.10 (20)    & 98.73 (18)    & 95.25 (18)   \\ \hline
    3    & 80.73 (17)  & 97.83 (20)    & 94.42 (19)    & -             & 89.55 (18)    & 95.09 (14)    & 89.08 (20)    & 91.50 (20)    & 74.18 (19)    & 84.58 (20)   \\ \hline
    4    & 72.20 (18)  & 96.18 (18)    & 96.09 (19)    & 89.55 (18)    & -             & 98.10 (16)    & 98.18 (18)    & 82.56 (19)    & 91.50 (20)    & 88.64 (20)   \\ \hline
    5    & 95.20 (16)  & 100.00 (3)    & 100.00 (9)    & 95.09 (14)    & 98.10 (16)    & -             & 94.18 (19)    & 100.00 (10)   & 95.90 (19)    & 98.45 (20)   \\ \hline
    6    & 92.36 (20)  & 100.00 (6)    & 99.92 (17)    & 89.08 (20)    & 98.18 (18)    & 94.18 (19)    & -             & 97.10 (20)    & 91.91 (18)    & 98.92 (20)   \\
\hline
    \end{tabular}
    \label{tab:qcnn_Qiskit}
\end{table*}

\begin{table*}[h]
    \centering
    \footnotesize
    \caption{HQNN4EOv2 - PennyLane - Avg Test Accuracy 92.51 and best model saved at epoch 16.11}
    \begin{tabular}{|c|c|c|c|c|c|c|c|c|c|c|}
    \hline
         & 0           & 1             & 2             & 3             & 4             & 5             & 6             & 7             & 8             & 9            \\ \hline
    0    & -           & 99.36 (16)    & 99.36 (18)    & 79.27 (19)    & 73.90 (20)    & 95.00 (20)    & 93.27 (20)    & 88.00 (20)    & 80.00 (20)    & 81.82 (20)   \\ \hline
    1    & 99.36 (16)  & -             & 88.50 (20)    & 97.92 (13)    & 96.36 (16)    & 100.00 (4)    & 100.00 (10)   & 92.30 (16)    & 99.36 (19)    & 97.67 (18)   \\ \hline
    2    & 99.36 (18)  & 88.50 (20)    & -             & 94.75 (20)    & 97.00 (20)    & 100.00 (6)    & 99.92 (15)    & 90.60 (19)    & 99.45 (20)    & 95.42 (19)   \\ \hline
    3    & 79.27 (19)  & 97.92 (13)    & 94.75 (20)    & -             & 89.45 (18)    & 95.09 (19)    & 88.42 (20)    & 91.20 (20)    & 74.55 (18)    & 84.42 (20)   \\ \hline
    4    & 73.90 (20)  & 96.36 (16)    & 97.00 (20)    & 89.45 (18)    & -             & 97.10 (14)    & 98.64 (20)    & 82.44 (20)    & 91.70 (18)    & 88.27 (19)   \\ \hline
    5    & 95.00 (20)  & 100.00 (4)    & 100.00 (6)    & 95.09 (19)    & 97.10 (14)    & -             & 94.55 (20)    & 100.00 (12)   & 95.90 (17)    & 98.55 (18)   \\ \hline
    6    & 93.27 (20)  & 100.00 (10)   & 99.92 (15)    & 88.42 (20)    & 98.64 (20)    & 94.55 (20)    & -             & 96.00 (17)    & 91.55 (19)    & 98.17 (18)   \\
\hline
    \end{tabular}
    \label{tab:qcnn_pennylane}
\end{table*}

\begin{table*}[h]
    \centering
    \footnotesize
    \caption{HQNN4EOv3 - Qiskit - Avg Test Accuracy 93.45 and best model saved at epoch 15.89}
    \begin{tabular}{|c|c|c|c|c|c|c|c|c|c|c|}
    \hline
         & 0           & 1             & 2             & 3             & 4             & 5             & 6             & 7             & 8             & 9            \\ \hline
    0    & -           & 99.00 (19)    & 98.09 (20)    & 79.45 (20)    & 75.90 (20)    & 94.30 (16)    & 90.18 (20)    & 92.56 (16)    & 80.30 (19)    & 89.45 (20)   \\ \hline
    1    & 99.00 (19)  & -             & 93.25 (20)    & 97.83 (19)    & 96.82 (18)    & 100.00 (6)    & 99.75 (17)    & 92.30 (10)    & 98.91 (13)    & 98.17 (20)   \\ \hline
    2    & 98.09 (20)  & 93.25 (20)    & -             & 95.75 (16)    & 96.55 (20)    & 99.82 (8)     & 99.50 (6)     & 87.40 (17)    & 99.09 (20)    & 95.83 (20)   \\ \hline
    3    & 79.45 (20)  & 97.83 (19)    & 95.75 (16)    & -             & 90.09 (20)    & 96.18 (18)    & 92.25 (18)    & 91.70 (17)    & 72.55 (20)    & 85.83 (19)   \\ \hline
    4    & 75.90 (20)  & 96.82 (18)    & 96.55 (20)    & 90.09 (20)    & -             & 97.90 (19)    & 96.55 (15)    & 90.78 (18)    & 92.20 (20)    & 91.64 (19)   \\ \hline
    5    & 94.30 (16)  & 100.00 (6)    & 99.82 (8)     & 96.18 (18)    & 97.90 (19)    & -             & 95.18 (19)    & 100.00 (4)    & 97.10 (16)    & 98.73 (11)   \\ \hline
    6    & 90.18 (20)  & 99.75 (17)    & 99.50 (6)     & 92.25 (18)    & 96.55 (15)    & 95.18 (19)    & -             & 98.70 (14)    & 95.18 (17)    & 98.42 (19)   \\
\hline
    \end{tabular}
    \label{tab:qnn4eo_Qiskit}
\end{table*}

\begin{table*}[ht]
    \centering
    \footnotesize
    \caption{HQNN4EOv3 - PennyLane - Avg Test Accuracy 93.15 and best model saved at epoch 15.46}
    \begin{tabular}{|c|c|c|c|c|c|c|c|c|c|c|}
    \hline
         & 0           & 1             & 2             & 3             & 4             & 5             & 6             & 7             & 8             & 9            \\ \hline
    0    & -           & 99.00 (20)    & 97.45 (11)    & 79.27 (17)    & 76.20 (18)    & 94.60 (17)    & 89.73 (18)    & 91.22 (10)    & 79.70 (17)    & 89.45 (18)   \\ \hline
    1    & 99.00 (20)  & -             & 92.17 (16)    & 98.08 (19)    & 96.09 (16)    & 100.00 (2)    & 99.83 (19)    & 92.10 (19)    & 98.55 (18)    & 98.17 (15)   \\ \hline
    2    & 97.45 (11)  & 92.17 (16)    & -             & 95.50 (19)    & 93.09 (19)    & 99.91 (8)     & 99.42 (7)     & 86.30 (20)    & 99.09 (18)    & 96.00 (20)   \\ \hline
    3    & 79.27 (17)  & 98.08 (19)    & 95.50 (19)    & -             & 89.91 (20)    & 96.09 (17)    & 92.42 (20)    & 92.00 (19)    & 71.55 (18)    & 83.67 (18)   \\ \hline
    4    & 76.20 (18)  & 96.09 (16)    & 93.09 (19)    & 89.91 (20)    & -             & 98.10 (17)    & 95.45 (17)    & 91.67 (20)    & 91.80 (20)    & 91.00 (17)   \\ \hline
    5    & 94.60 (17)  & 100.00 (2)    & 99.91 (8)     & 96.09 (17)    & 98.10 (17)    & -             & 95.09 (16)    & 99.89 (4)     & 97.30 (17)    & 99.45 (18)   \\ \hline
    6    & 89.73 (18)  & 99.83 (19)    & 99.42 (7)     & 92.42 (20)    & 95.45 (17)    & 95.09 (16)    & -             & 99.20 (20)    & 96.45 (17)    & 98.83 (11)   \\
\hline
    \end{tabular}
    \label{tab:qnn4eo_pennylane}
\end{table*}

\begin{table*}[h]
    \centering
    \footnotesize
    \caption{HQViT - Qiskit - Avg Test Accuracy 87.95 and best model saved at epoch 16.25}
    \begin{tabular}{|c|c|c|c|c|c|c|c|c|c|c|}
    \hline
         & 0           & 1             & 2             & 3             & 4             & 5             & 6             & 7             & 8             & 9            \\ \hline
    0    & -           & 95.00 (17)    & 90.64 (20)    & 78.73 (18)    & 69.00 (19)    & 85.30 (19)    & 78.09 (20)    & 80.00 (19)    & 75.00 (16)    & 77.36 (20)   \\ \hline
    1    & 95.00 (17)  & -             & 87.75 (20)    & 97.83 (13)    & 88.45 (15)    & 99.55 (6)     & 99.08 (15)    & 90.70 (19)    & 97.64 (15)    & 97.25 (10)   \\ \hline
    2    & 90.64 (20)  & 87.75 (20)    & -             & 94.17 (18)    & 84.09 (18)    & 98.91 (17)    & 95.33 (14)    & 77.80 (15)    & 94.45 (18)    & 94.17 (20)   \\ \hline
    3    & 78.73 (18)  & 97.83 (13)    & 94.17 (18)    & -             & 87.82 (19)    & 94.09 (17)    & 76.83 (19)    & 92.50 (20)    & 70.27 (18)    & 82.42 (20)   \\ \hline
    4    & 69.00 (19)  & 88.45 (15)    & 84.09 (18)    & 87.82 (19)    & -             & 93.70 (19)    & 88.64 (19)    & 85.78 (19)    & 91.00 (18)    & 87.27 (14)   \\ \hline
    5    & 85.30 (19)  & 99.55 (6)     & 98.91 (17)    & 94.09 (17)    & 93.70 (19)    & -             & 88.18 (19)    & 97.78 (20)    & 95.00 (17)    & 96.91 (15)   \\ \hline
    6    & 78.09 (20)  & 99.08 (15)    & 95.33 (14)    & 76.83 (19)    & 88.64 (19)    & 88.18 (19)    & -             & 93.90 (16)    & 84.45 (18)    & 82.25 (20)   \\
\hline
    \end{tabular}
    \label{tab:qvit_Qiskit}
\end{table*}

\begin{table*}[h]
    \centering
    \footnotesize
    \caption{HQViT - PennyLane - Avg Test Accuracy 87.77 and best model saved at epoch 16.20}
    \begin{tabular}{|c|c|c|c|c|c|c|c|c|c|c|}
    \hline
         & 0           & 1             & 2             & 3             & 4             & 5             & 6             & 7             & 8             & 9            \\ \hline
    0    & -           & 95.09 (20)    & 91.09 (19)    & 76.82 (15)    & 69.20 (20)    & 88.10 (20)    & 78.27 (20)    & 86.44 (20)    & 75.10 (19)    & 76.36 (16)   \\ \hline
    1    & 95.09 (20)  & -             & 82.42 (19)    & 98.17 (20)    & 88.64 (15)    & 99.64 (5)     & 98.58 (10)    & 84.80 (16)    & 97.91 (6)     & 97.67 (15)   \\ \hline
    2    & 91.09 (19)  & 82.42 (19)    & -             & 94.33 (14)    & 85.73 (19)    & 99.73 (20)    & 95.33 (8)     & 78.20 (17)    & 94.55 (20)    & 93.83 (18)   \\ \hline
    3    & 76.82 (15)  & 98.17 (20)    & 94.33 (14)    & -             & 88.09 (20)    & 94.09 (17)    & 84.25 (20)    & 91.80 (19)    & 69.55 (18)    & 80.67 (19)   \\ \hline
    4    & 69.20 (20)  & 88.64 (15)    & 85.73 (19)    & 88.09 (20)    & -             & 94.00 (20)    & 90.09 (15)    & 89.22 (20)    & 91.10 (19)    & 85.00 (16)   \\ \hline
    5    & 88.10 (20)  & 99.64 (5)     & 99.73 (20)    & 94.09 (17)    & 94.00 (20)    & -             & 87.82 (20)    & 96.89 (18)    & 94.70 (18)    & 96.45 (18)   \\ \hline
    6    & 78.27 (20)  & 98.58 (10)    & 95.33 (8)     & 84.25 (20)    & 90.09 (15)    & 87.82 (20)    & -             & 92.40 (18)    & 74.18 (19)    & 81.25 (15)   \\
\hline
    \end{tabular}
    \label{tab:qvit_pennylane}
\end{table*}

\begin{table*}[h]
    \centering
    \caption{NN4EOv1 - mean (max, min) values with 10 seeds}
    \resizebox{\textwidth}{!}{%
    \begin{tabular}{|c|c|c|c|c|c|c|c|c|c|c|}
        \hline
         & \multirow{2}{*}{0} & \multirow{2}{*}{1} & \multirow{2}{*}{2} & \multirow{2}{*}{3} & \multirow{2}{*}{4} & \multirow{2}{*}{5} & \multirow{2}{*}{6} & \multirow{2}{*}{7} & \multirow{2}{*}{8} & \multirow{2}{*}{9} \\
         &  &  &  &  &  &  &  &  &  &  \\
        \hline
        \multirow{2}{*}{0} & \multirow{2}{*}{-} & 95.43 & 93.04 & 78.67 & 72.34 & 93.71 & 87.15 & 89.23 & 77.17 & 85.45 \\
        & & (97.18, 94.45) & (98.45, 87.82) & (82.73, 72.91) & (75.00, 68.50) & (94.70, 92.10) & (88.82, 85.36) & (91.89, 85.56) & (79.40, 73.70) & (87.73, 82.91) \\
        \hline
        \multirow{2}{*}{1} & 95.43 & \multirow{2}{*}{-} & 89.81 & 97.41 & 91.20 & 99.94 & 99.13 & 92.40 & 98.75 & 97.88 \\
        & (97.18, 94.45) & & (92.33, 86.67) & (97.92, 96.92) & (92.09, 89.82) & (100.00, 99.82) & (99.75, 98.50) & (93.40, 91.00) & (99.27, 97.73) & (98.50, 97.42) \\
        \hline
        \multirow{2}{*}{2} & 93.04 & 89.81 & \multirow{2}{*}{-} & 92.49 & 86.32 & 99.74 & 98.47 & 86.70 & 93.27 & 92.30 \\
        & (98.45, 87.82) & (92.33, 86.67) & & (94.75, 88.75) & (97.09, 73.18) & (100.00, 99.18) & (99.92, 96.33) & (91.60, 75.30) & (95.82, 88.18) & (93.42, 90.83) \\
\hline
    \end{tabular}%
    }
    \label{tab:accuracy_matrix_nn4eov1}
\end{table*}

\begin{table*}[h]
    \centering
    \caption{HQNN4EOv1 - mean (max, min) values with 10 seeds}
    \resizebox{\textwidth}{!}{%
    \begin{tabular}{|c|c|c|c|c|c|c|c|c|c|c|}
        \hline
         & \multirow{2}{*}{0} & \multirow{2}{*}{1} & \multirow{2}{*}{2} & \multirow{2}{*}{3} & \multirow{2}{*}{4} & \multirow{2}{*}{5} & \multirow{2}{*}{6} & \multirow{2}{*}{7} & \multirow{2}{*}{8} & \multirow{2}{*}{9} \\
         &  &  &  &  &  &  &  &  &  &  \\
        \hline
        \multirow{2}{*}{0} & \multirow{2}{*}{-} & 93.73 & 94.13 & 77.82 & 72.93 & 93.32 & 87.50 & 89.62 & 76.24 & 86.42 \\
        & & (95.73, 86.73) & (97.09, 83.55) & (81.45, 72.73) & (74.70, 70.70) & (94.90, 90.90) & (90.36, 85.45) & (91.67, 87.11) & (78.00, 73.60) & (90.45, 83.55) \\
        \hline
        \multirow{2}{*}{1} & 93.73 & \multirow{2}{*}{-} & 88.17 & 97.32 & 88.22 & 99.85 & 98.72 & 92.56 & 98.49 & 96.59 \\
        & (95.73, 86.73) & & (92.92, 84.08) & (98.00, 95.83) & (91.27, 86.27) & (100.00, 99.45) & (99.67, 96.00) & (94.00, 91.30) & (99.36, 97.55) & (97.58, 95.75) \\
        \hline
        \multirow{2}{*}{2} & 94.13 & 88.17 & \multirow{2}{*}{-} & 93.79 & 86.64 & 99.81 & 98.77 & 89.78 & 95.33 & 91.61 \\
        & (97.09, 83.55) & (92.92, 84.08) & & (95.58, 92.17) & (95.09, 80.45) & (100.00, 99.45) & (99.50, 96.67) & (96.10, 84.90) & (99.00, 89.55) & (93.00, 89.83) \\
\hline
    \end{tabular}%
    }
    \label{tab:accuracy_matrix_hqnn4eo_v1}
\end{table*}

\begin{table*}[h]
    \centering
    \caption{NN4EOv2 - mean (max, min) values with 10 seeds}
    \resizebox{\textwidth}{!}{%
    \begin{tabular}{|c|c|c|c|c|c|c|c|c|c|c|}
        \hline
         & \multirow{2}{*}{0} & \multirow{2}{*}{1} & \multirow{2}{*}{2} & \multirow{2}{*}{3} & \multirow{2}{*}{4} & \multirow{2}{*}{5} & \multirow{2}{*}{6} & \multirow{2}{*}{7} & \multirow{2}{*}{8} & \multirow{2}{*}{9} \\
        & & & & & & & & & & \\
        \hline
        \multirow{2}{*}{0} & \multirow{2}{*}{-} & \multicolumn{1}{c|}{98.49} & \multicolumn{1}{c|}{98.06} & \multicolumn{1}{c|}{80.74} & \multicolumn{1}{c|}{71.37} & \multicolumn{1}{c|}{94.33} & \multicolumn{1}{c|}{88.81} & \multicolumn{1}{c|}{90.02} & \multicolumn{1}{c|}{77.46} & \multicolumn{1}{c|}{84.90} \\
        & & \multicolumn{1}{c|}{(99.27, 97.18)} & \multicolumn{1}{c|}{(99.36, 96.27)} & \multicolumn{1}{c|}{(83.82, 77.82)} & \multicolumn{1}{c|}{(75.2, 69.1)} & \multicolumn{1}{c|}{(95.9, 93.2)} & \multicolumn{1}{c|}{(96.18, 84.91)} & \multicolumn{1}{c|}{(93.44, 83.22)} & \multicolumn{1}{c|}{(79.7, 72.0)} & \multicolumn{1}{c|}{(88.27, 79.64)} \\
        \hline
        \multirow{2}{*}{1} & \multicolumn{1}{c|}{98.49} & \multirow{2}{*}{-} & \multicolumn{1}{c|}{84.62} & \multicolumn{1}{c|}{97.38} & \multicolumn{1}{c|}{94.71} & \multicolumn{1}{c|}{99.98} & \multicolumn{1}{c|}{99.47} & \multicolumn{1}{c|}{92.52} & \multicolumn{1}{c|}{98.87} & \multicolumn{1}{c|}{98.04} \\
        & \multicolumn{1}{c|}{(99.27, 97.18)} & & \multicolumn{1}{c|}{(93.75, 74.08)} & \multicolumn{1}{c|}{(98.25, 96.58)} & \multicolumn{1}{c|}{(96.73, 92.36)} & \multicolumn{1}{c|}{(100.0, 99.82)} & \multicolumn{1}{c|}{(99.92, 99.08)} & \multicolumn{1}{c|}{(94.2, 91.5)} & \multicolumn{1}{c|}{(99.64, 97.45)} & \multicolumn{1}{c|}{(98.58, 97.5)} \\
        \hline
        \multirow{2}{*}{2} & \multicolumn{1}{c|}{98.06} & \multicolumn{1}{c|}{84.62} & \multirow{2}{*}{-} & \multicolumn{1}{c|}{93.95} & \multicolumn{1}{c|}{95.39} & \multicolumn{1}{c|}{99.65} & \multicolumn{1}{c|}{99.75} & \multicolumn{1}{c|}{91.08} & \multicolumn{1}{c|}{97.11} & \multicolumn{1}{c|}{94.76} \\
\hline
    \end{tabular}%
    }
    \label{tab:accuracy_matrix_cnn}
\end{table*}

\begin{table*}[h]
    \centering
    \caption{HQNN4EOv2 - mean (max, min) values with 10 seeds}
    \resizebox{\textwidth}{!}{%
    \begin{tabular}{|c|c|c|c|c|c|c|c|c|c|c|}
        \hline
         & \multirow{2}{*}{0} & \multirow{2}{*}{1} & \multirow{2}{*}{2} & \multirow{2}{*}{3} & \multirow{2}{*}{4} & \multirow{2}{*}{5} & \multirow{2}{*}{6} & \multirow{2}{*}{7} & \multirow{2}{*}{8} & \multirow{2}{*}{9} \\
        & & & & & & & & & & \\
        \hline
        \multirow{2}{*}{0} & \multirow{2}{*}{-} & \multicolumn{1}{c|}{97.91} & \multicolumn{1}{c|}{98.07} & \multicolumn{1}{c|}{80.71} & \multicolumn{1}{c|}{71.82} & \multicolumn{1}{c|}{94.35} & \multicolumn{1}{c|}{88.50} & \multicolumn{1}{c|}{91.03} & \multicolumn{1}{c|}{77.05} & \multicolumn{1}{c|}{85.35} \\
        & & \multicolumn{1}{c|}{(98.91, 94.45)} & \multicolumn{1}{c|}{(99.55, 91.09)} & \multicolumn{1}{c|}{(83.55, 76.55)} & \multicolumn{1}{c|}{(75.7, 67.6)} & \multicolumn{1}{c|}{(96.0, 93.1)} & \multicolumn{1}{c|}{(93.82, 85.18)} & \multicolumn{1}{c|}{(94.11, 87.56)} & \multicolumn{1}{c|}{(80.0, 72.3)} & \multicolumn{1}{c|}{(89.45, 78.0)} \\
        \hline
        \multirow{2}{*}{1} & \multicolumn{1}{c|}{97.91} & \multirow{2}{*}{-} & \multicolumn{1}{c|}{87.60} & \multicolumn{1}{c|}{97.50} & \multicolumn{1}{c|}{95.00} & \multicolumn{1}{c|}{99.98} & \multicolumn{1}{c|}{99.61} & \multicolumn{1}{c|}{92.60} & \multicolumn{1}{c|}{98.75} & \multicolumn{1}{c|}{97.97} \\
        & \multicolumn{1}{c|}{(98.91, 94.45)} & & \multicolumn{1}{c|}{(95.83, 79.08)} & \multicolumn{1}{c|}{(98.42, 96.67)} & \multicolumn{1}{c|}{(97.09, 91.73)} & \multicolumn{1}{c|}{(100.0, 99.82)} & \multicolumn{1}{c|}{(100.0, 99.0)} & \multicolumn{1}{c|}{(93.6, 91.4)} & \multicolumn{1}{c|}{(99.55, 97.64)} & \multicolumn{1}{c|}{(98.58, 95.92)} \\
        \hline
        \multirow{2}{*}{2} & \multicolumn{1}{c|}{98.07} & \multicolumn{1}{c|}{87.60} & \multirow{2}{*}{-} & \multicolumn{1}{c|}{94.29} & \multicolumn{1}{c|}{96.03} & \multicolumn{1}{c|}{99.92} & \multicolumn{1}{c|}{99.87} & \multicolumn{1}{c|}{88.82} & \multicolumn{1}{c|}{98.50} & \multicolumn{1}{c|}{94.67} \\
\hline
    \end{tabular}%
    }
    \label{tab:accuracy_matrix_qcnn}
\end{table*}

\begin{table*}[h]
    \centering
    \caption{NN4EOv3 - mean (max, min) values with 10 seeds}
    \resizebox{\textwidth}{!}{%
    \begin{tabular}{|c|c|c|c|c|c|c|c|c|c|c|}
        \hline
         & \multirow{2}{*}{0} & \multirow{2}{*}{1} & \multirow{2}{*}{2} & \multirow{2}{*}{3} & \multirow{2}{*}{4} & \multirow{2}{*}{5} & \multirow{2}{*}{6} & \multirow{2}{*}{7} & \multirow{2}{*}{8} & \multirow{2}{*}{9} \\
        & & & & & & & & & & \\
        \hline
        \multirow{2}{*}{0} & \multirow{2}{*}{-} & \multicolumn{1}{c|}{98.11} & \multicolumn{1}{c|}{98.40} & \multicolumn{1}{c|}{81.45} & \multicolumn{1}{c|}{74.45} & \multicolumn{1}{c|}{94.50} & \multicolumn{1}{c|}{88.71} & \multicolumn{1}{c|}{91.74} & \multicolumn{1}{c|}{78.64} & \multicolumn{1}{c|}{87.72} \\
        & & \multicolumn{1}{c|}{(98.73, 97.27)} & \multicolumn{1}{c|}{(99.45, 97.09)} & \multicolumn{1}{c|}{(84.36, 77.82)} & \multicolumn{1}{c|}{(77.4, 71.3)} & \multicolumn{1}{c|}{(95.4, 93.6)} & \multicolumn{1}{c|}{(92.45, 83.18)} & \multicolumn{1}{c|}{(93.78, 89.56)} & \multicolumn{1}{c|}{(81.6, 75.6)} & \multicolumn{1}{c|}{(90.27, 84.18)} \\
        \hline
        \multirow{2}{*}{1} & \multicolumn{1}{c|}{98.11} & \multirow{2}{*}{-} & \multicolumn{1}{c|}{90.22} & \multicolumn{1}{c|}{97.54} & \multicolumn{1}{c|}{94.51} & \multicolumn{1}{c|}{99.93} & \multicolumn{1}{c|}{99.34} & \multicolumn{1}{c|}{92.32} & \multicolumn{1}{c|}{98.82} & \multicolumn{1}{c|}{98.23} \\
        & \multicolumn{1}{c|}{(98.73, 97.27)} & & \multicolumn{1}{c|}{(94.0, 83.5)} & \multicolumn{1}{c|}{(98.08, 97.0)} & \multicolumn{1}{c|}{(96.36, 91.73)} & \multicolumn{1}{c|}{(100.0, 99.73)} & \multicolumn{1}{c|}{(99.83, 98.5)} & \multicolumn{1}{c|}{(93.4, 91.0)} & \multicolumn{1}{c|}{(99.73, 97.91)} & \multicolumn{1}{c|}{(99.0, 97.58)} \\
        \hline
        \multirow{2}{*}{2} & \multicolumn{1}{c|}{98.40} & \multicolumn{1}{c|}{90.22} & \multirow{2}{*}{-} & \multicolumn{1}{c|}{95.46} & \multicolumn{1}{c|}{96.18} & \multicolumn{1}{c|}{99.89} & \multicolumn{1}{c|}{99.67} & \multicolumn{1}{c|}{90.44} & \multicolumn{1}{c|}{97.92} & \multicolumn{1}{c|}{95.39} \\
\hline
    \end{tabular}%
    }
    \label{tab:accuracy_matrix_nn4eo}
\end{table*}

\begin{table*}[h]
    \centering
    \caption{HQNN4EOv3 - mean (max, min) values with 10 seeds}
    \resizebox{\textwidth}{!}{%
    \begin{tabular}{|c|c|c|c|c|c|c|c|c|c|c|}
        \hline
         & \multirow{2}{*}{0} & \multirow{2}{*}{1} & \multirow{2}{*}{2} & \multirow{2}{*}{3} & \multirow{2}{*}{4} & \multirow{2}{*}{5} & \multirow{2}{*}{6} & \multirow{2}{*}{7} & \multirow{2}{*}{8} & \multirow{2}{*}{9} \\
        & & & & & & & & & & \\
        \hline
        \multirow{2}{*}{0} & \multirow{2}{*}{-} & \multicolumn{1}{c|}{98.23} & \multicolumn{1}{c|}{97.99} & \multicolumn{1}{c|}{81.28} & \multicolumn{1}{c|}{74.60} & \multicolumn{1}{c|}{94.30} & \multicolumn{1}{c|}{88.26} & \multicolumn{1}{c|}{91.91} & \multicolumn{1}{c|}{78.04} & \multicolumn{1}{c|}{86.68} \\
        & & \multicolumn{1}{c|}{(98.91, 97.55)} & \multicolumn{1}{c|}{(99.36, 93.82)} & \multicolumn{1}{c|}{(83.55, 79.09)} & \multicolumn{1}{c|}{(76.5, 71.5)} & \multicolumn{1}{c|}{(95.3, 92.6)} & \multicolumn{1}{c|}{(90.91, 80.73)} & \multicolumn{1}{c|}{(93.78, 90.44)} & \multicolumn{1}{c|}{(81.4, 74.6)} & \multicolumn{1}{c|}{(90.45, 82.45)} \\
        \hline
        \multirow{2}{*}{1} & \multicolumn{1}{c|}{98.23} & \multirow{2}{*}{-} & \multicolumn{1}{c|}{91.55} & \multicolumn{1}{c|}{97.75} & \multicolumn{1}{c|}{94.92} & \multicolumn{1}{c|}{99.98} & \multicolumn{1}{c|}{99.42} & \multicolumn{1}{c|}{92.64} & \multicolumn{1}{c|}{98.91} & \multicolumn{1}{c|}{98.26} \\
        & \multicolumn{1}{c|}{(98.91, 97.55)} & & \multicolumn{1}{c|}{(94.58, 87.92)} & \multicolumn{1}{c|}{(98.33, 97.17)} & \multicolumn{1}{c|}{(96.55, 92.27)} & \multicolumn{1}{c|}{(100.0, 99.82)} & \multicolumn{1}{c|}{(99.83, 98.58)} & \multicolumn{1}{c|}{(94.1, 91.5)} & \multicolumn{1}{c|}{(99.64, 97.64)} & \multicolumn{1}{c|}{(99.17, 97.58)} \\
        \hline
        \multirow{2}{*}{2} & \multicolumn{1}{c|}{97.99} & \multicolumn{1}{c|}{91.55} & \multirow{2}{*}{-} & \multicolumn{1}{c|}{95.18} & \multicolumn{1}{c|}{96.30} & \multicolumn{1}{c|}{99.91} & \multicolumn{1}{c|}{99.51} & \multicolumn{1}{c|}{92.65} & \multicolumn{1}{c|}{98.32} & \multicolumn{1}{c|}{95.09} \\
\hline
    \end{tabular}%
    }
    \label{tab:accuracy_matrix_qnn4eo}
\end{table*}

\begin{table*}[h]
    \centering
    \caption{VIT - mean (max, min) values with 10 seeds}
    \resizebox{\textwidth}{!}{%
    \begin{tabular}{|c|c|c|c|c|c|c|c|c|c|c|}
        \hline
         & \multirow{2}{*}{0} & \multirow{2}{*}{1} & \multirow{2}{*}{2} & \multirow{2}{*}{3} & \multirow{2}{*}{4} & \multirow{2}{*}{5} & \multirow{2}{*}{6} & \multirow{2}{*}{7} & \multirow{2}{*}{8} & \multirow{2}{*}{9} \\
        & & & & & & & & & & \\
        \hline
        \multirow{2}{*}{0} & \multirow{2}{*}{-} & \multicolumn{1}{c|}{95.30} & \multicolumn{1}{c|}{89.76} & \multicolumn{1}{c|}{77.10} & \multicolumn{1}{c|}{69.17} & \multicolumn{1}{c|}{88.11} & \multicolumn{1}{c|}{75.78} & \multicolumn{1}{c|}{82.67} & \multicolumn{1}{c|}{76.13} & \multicolumn{1}{c|}{77.06} \\
        & & \multicolumn{1}{c|}{(96.18, 93.91)} & \multicolumn{1}{c|}{(94.64, 83.64)} & \multicolumn{1}{c|}{(80.73, 67.27)} & \multicolumn{1}{c|}{(71.1, 66.1)} & \multicolumn{1}{c|}{(92.0, 84.4)} & \multicolumn{1}{c|}{(78.55, 72.36)} & \multicolumn{1}{c|}{(88.33, 77.33)} & \multicolumn{1}{c|}{(78.3, 73.5)} & \multicolumn{1}{c|}{(79.82, 74.73)} \\
        \hline
        \multirow{2}{*}{1} & \multicolumn{1}{c|}{95.30} & \multirow{2}{*}{-} & \multicolumn{1}{c|}{85.56} & \multicolumn{1}{c|}{97.33} & \multicolumn{1}{c|}{89.74} & \multicolumn{1}{c|}{99.56} & \multicolumn{1}{c|}{98.74} & \multicolumn{1}{c|}{91.84} & \multicolumn{1}{c|}{98.62} & \multicolumn{1}{c|}{97.74} \\
        & \multicolumn{1}{c|}{(96.18, 93.91)} & & \multicolumn{1}{c|}{(89.08, 80.5)} & \multicolumn{1}{c|}{(98.08, 96.25)} & \multicolumn{1}{c|}{(91.0, 88.73)} & \multicolumn{1}{c|}{(100.0, 98.73)} & \multicolumn{1}{c|}{(99.5, 97.0)} & \multicolumn{1}{c|}{(93.5, 88.9)} & \multicolumn{1}{c|}{(99.36, 97.91)} & \multicolumn{1}{c|}{(98.5, 96.92)} \\
        \hline
        \multirow{2}{*}{2} & \multicolumn{1}{c|}{89.76} & \multicolumn{1}{c|}{85.56} & \multirow{2}{*}{-} & \multicolumn{1}{c|}{93.09} & \multicolumn{1}{c|}{82.72} & \multicolumn{1}{c|}{99.12} & \multicolumn{1}{c|}{95.34} & \multicolumn{1}{c|}{77.74} & \multicolumn{1}{c|}{93.62} & \multicolumn{1}{c|}{93.20} \\
\hline
    \end{tabular}%
    }
    \label{tab:accuracy_matrix_vit}
\end{table*}

\begin{table*}[h]
    \centering
    \caption{HQVIT - mean (max, min) values with 10 seeds}
    \resizebox{\textwidth}{!}{%
    \begin{tabular}{|c|c|c|c|c|c|c|c|c|c|c|}
        \hline
         & \multirow{2}{*}{0} & \multirow{2}{*}{1} & \multirow{2}{*}{2} & \multirow{2}{*}{3} & \multirow{2}{*}{4} & \multirow{2}{*}{5} & \multirow{2}{*}{6} & \multirow{2}{*}{7} & \multirow{2}{*}{8} & \multirow{2}{*}{9} \\
        & & & & & & & & & & \\
        \hline
        \multirow{2}{*}{0} & \multirow{2}{*}{-} & \multicolumn{1}{c|}{94.23} & \multicolumn{1}{c|}{89.72} & \multicolumn{1}{c|}{78.40} & \multicolumn{1}{c|}{69.15} & \multicolumn{1}{c|}{85.98} & \multicolumn{1}{c|}{76.69} & \multicolumn{1}{c|}{84.75} & \multicolumn{1}{c|}{75.87} & \multicolumn{1}{c|}{77.43} \\
        & & \multicolumn{1}{c|}{(95.82, 91.64)} & \multicolumn{1}{c|}{(95.00, 85.18)} & \multicolumn{1}{c|}{(81.91, 75.36)} & \multicolumn{1}{c|}{(70.4, 66.8)} & \multicolumn{1}{c|}{(89.4, 78.9)} & \multicolumn{1}{c|}{(78.27, 74.73)} & \multicolumn{1}{c|}{(89.56, 77.67)} & \multicolumn{1}{c|}{(77.8, 72.3)} & \multicolumn{1}{c|}{(79.73, 75.18)} \\
        \hline
        \multirow{2}{*}{1} & \multicolumn{1}{c|}{94.23} & \multirow{2}{*}{-} & \multicolumn{1}{c|}{84.90} & \multicolumn{1}{c|}{97.30} & \multicolumn{1}{c|}{88.56} & \multicolumn{1}{c|}{99.59} & \multicolumn{1}{c|}{98.61} & \multicolumn{1}{c|}{90.97} & \multicolumn{1}{c|}{98.65} & \multicolumn{1}{c|}{97.61} \\
        & \multicolumn{1}{c|}{(95.82, 91.64)} & & \multicolumn{1}{c|}{(87.83, 78.42)} & \multicolumn{1}{c|}{(98.25, 95.58)} & \multicolumn{1}{c|}{(90.45, 84.0)} & \multicolumn{1}{c|}{(100.0, 98.73)} & \multicolumn{1}{c|}{(99.58, 94.33)} & \multicolumn{1}{c|}{(92.4, 86.6)} & \multicolumn{1}{c|}{(99.0, 97.91)} & \multicolumn{1}{c|}{(98.42, 96.83)} \\
        \hline
        \multirow{2}{*}{2} & \multicolumn{1}{c|}{89.72} & \multicolumn{1}{c|}{84.90} & \multirow{2}{*}{-} & \multicolumn{1}{c|}{92.76} & \multicolumn{1}{c|}{82.32} & \multicolumn{1}{c|}{97.66} & \multicolumn{1}{c|}{94.66} & \multicolumn{1}{c|}{78.80} & \multicolumn{1}{c|}{93.79} & \multicolumn{1}{c|}{92.93} \\
\hline
    \end{tabular}%
    }
    \label{tab:accuracy_matrix_qvit}
\end{table*}

\begin{table*}[h]
    \centering
    \caption{NN4EOv1 - Variance of Test Accuracy across 10 seed}
    \begin{tabular}{|c | c | c | c | c | c | c | c | c | c|}
    \hline
    0 & 1 & 2 & 3 & 4 & 5 & 6 & 7 & 8 & 9 \\
    \hline
    - & 0.58 & 9.65 & 8.22 & 2.86 & 0.50 & 1.20 & 4.86 & 2.34 & 2.66 \\
    \hline
    0.58 & - & 3.76 & 0.11 & 0.51 & 0.00 & 0.13 & 0.39 & 0.20 & 0.13 \\
    \hline
    9.65 & 3.76 & - & 3.05 & 41.19 & 0.05 & 1.55 & 22.56 & 4.93 & 0.40 \\
    \hline
    8.22 & 0.11 & 3.05 & - & 1.45 & 0.50 & 2.24 & 1.43 & 5.55 & 15.84 \\
    \hline
    2.86 & 0.51 & 41.19 & 1.45 & - & 0.78 & 5.59 & 7.95 & 1.86 & 1.40 \\
    \hline
    0.50 & 0.00 & 0.05 & 0.50 & 0.78 & - & 2.37 & 0.16 & 0.29 & 0.05 \\
    \hline
    1.20 & 0.13 & 1.55 & 2.24 & 5.59 & 2.37 & - & 1.58 & 1.07 & 0.25 \\
    \hline
    4.86 & 0.39 & 22.56 & 1.43 & 7.95 & 0.16 & 1.58 & - & 0.83 & 0.88 \\
    \hline
    2.34 & 0.20 & 4.93 & 5.55 & 1.86 & 0.29 & 1.07 & 0.83 & - & 9.27 \\
    \hline
    2.66 & 0.13 & 0.40 & 15.84 & 1.40 & 0.05 & 0.25 & 0.88 & 9.27 & - \\
    \hline
    \end{tabular}
    \label{tab:variance_accuracy_nn4eov1}
\end{table*}

\begin{table*}[h]
    \centering
    \caption{HQNN4EOv1 - Variance of Test Accuracy across 10 seed}
    \begin{tabular}{|c | c | c | c | c | c | c | c | c | c|}
    \hline
    0 & 1 & 2 & 3 & 4 & 5 & 6 & 7 & 8 & 9 \\
    \hline
    - & 9.21 & 13.87 & 6.31 & 1.35 & 1.41 & 1.98 & 1.64 & 1.90 & 3.95 \\
    \hline
    9.21 & - & 7.09 & 0.39 & 2.92 & 0.02 & 1.08 & 0.63 & 0.29 & 0.31 \\
    \hline
    13.87 & 7.09 & - & 1.27 & 24.59 & 0.03 & 1.05 & 8.49 & 9.55 & 0.67 \\
    \hline
    6.31 & 0.39 & 1.27 & - & 0.95 & 0.48 & 2.58 & 1.19 & 6.26 & 11.12 \\
    \hline
    1.35 & 2.92 & 24.59 & 0.95 & - & 0.82 & 0.79 & 5.40 & 0.52 & 1.22 \\
    \hline
    1.41 & 0.02 & 0.03 & 0.48 & 0.82 & - & 1.84 & 0.18 & 0.41 & 0.14 \\
    \hline
    1.98 & 1.08 & 1.05 & 2.58 & 0.79 & 1.84 & - & 1.42 & 1.01 & 0.27 \\
    \hline
    1.64 & 0.63 & 8.49 & 1.19 & 5.40 & 0.18 & 1.42 & - & 1.55 & 0.26 \\
    \hline
    1.90 & 0.29 & 9.55 & 6.26 & 0.52 & 0.41 & 1.01 & 1.55 & - & 7.78 \\
    \hline
    3.95 & 0.31 & 0.67 & 11.12 & 1.22 & 0.14 & 0.27 & 0.26 & 7.78 & - \\
    \hline
    \end{tabular}
    \label{tab:variance_accuracy_hqnn4eov1}
\end{table*}

\begin{table*}[h]
    \centering
    \caption{NN4EOv2 - Variance of Test Accuracy across 10 seed}
    \begin{tabular}{|c | c | c | c | c | c | c | c | c | c|}
        \hline
        0 & 1 & 2 & 3 & 4 & 5 & 6 & 7 & 8 & 9 \\
        \hline
        - & 0.33 & 0.68 & 3.68 & 3.85 & 0.48 & 12.03 & 7.70 & 4.96 & 8.16 \\
        \hline
        0.33 & - & 35.97 & 0.31 & 0.98 & 0.00 & 0.05 & 0.49 & 0.12 & 0.15 \\
        \hline
        0.68 & 35.97 & - & 1.95 & 3.65 & 0.31 & 0.05 & 4.94 & 1.10 & 0.39 \\
        \hline
        3.68 & 0.31 & 1.95 & - & 4.75 & 0.22 & 1.70 & 7.24 & 8.80 & 11.79 \\
        \hline
        3.85 & 0.98 & 3.65 & 4.75 & - & 0.23 & 2.06 & 30.40 & 0.65 & 4.24 \\
        \hline
        0.48 & 0.00 & 0.31 & 0.22 & 0.23 & - & 2.88 & 0.10 & 0.36 & 0.67 \\
        \hline
        12.03 & 0.05 & 0.05 & 1.70 & 2.06 & 2.88 & - & 3.26 & 1.22 & 0.68 \\
        \hline
        7.70 & 0.49 & 4.94 & 7.24 & 30.40 & 0.10 & 3.26 & - & 2.41 & 0.36 \\
        \hline
        4.96 & 0.12 & 1.10 & 8.80 & 0.65 & 0.36 & 1.22 & 2.41 & - & 15.07 \\
        \hline
        8.16 & 0.15 & 0.39 & 11.79 & 4.24 & 0.67 & 0.68 & 0.36 & 15.07 & - \\
        \hline
    \end{tabular}
    \label{tab:variance_accuracy_cnn}
\end{table*}

\begin{table*}[h]
    \centering
    \caption{HQNN4EOv2 - Variance of Test Accuracy across 10 seed}
    \begin{tabular}{|c | c | c | c | c | c | c | c | c | c|}
        \hline
        0 & 1 & 2 & 3 & 4 & 5 & 6 & 7 & 8 & 9 \\
        \hline
        - & 1.67 & 5.64 & 4.82 & 4.30 & 0.49 & 4.69 & 5.16 & 5.15 & 12.59 \\
        \hline
        1.67 & - & 26.22 & 0.23 & 0.74 & 0.00 & 0.05 & 0.54 & 0.22 & 0.57 \\
        \hline
        5.64 & 26.22 & - & 6.38 & 1.78 & 0.01 & 0.03 & 28.59 & 1.14 & 1.48 \\
        \hline
        4.82 & 0.23 & 6.38 & - & 2.80 & 0.26 & 2.48 & 5.90 & 5.26 & 14.57 \\
        \hline
        4.30 & 0.74 & 1.78 & 2.80 & - & 0.52 & 3.84 & 16.96 & 0.62 & 1.88 \\
        \hline
        0.49 & 0.00 & 0.01 & 0.26 & 0.52 & - & 2.14 & 0.02 & 0.16 & 0.04 \\
        \hline
        4.69 & 0.05 & 0.03 & 2.48 & 3.84 & 2.14 & - & 0.86 & 0.87 & 0.35 \\
        \hline
        5.16 & 0.54 & 28.59 & 5.90 & 16.96 & 0.02 & 0.86 & - & 3.25 & 0.60 \\
        \hline
        5.15 & 0.22 & 1.14 & 5.26 & 0.62 & 0.16 & 0.87 & 3.25 & - & 23.79 \\
        \hline
        12.59 & 0.57 & 1.48 & 14.57 & 1.88 & 0.04 & 0.35 & 0.60 & 23.79 & - \\
        \hline
    \end{tabular}
    \label{tab:variance_accuracy_qcnn}
\end{table*}

\begin{table*}[h]
    \centering
    \caption{NN4EOv3 - Variance of Test Accuracy across 10 seed}
    \begin{tabular}{|c | c | c | c | c | c | c | c | c | c|}
        \hline
        0 & 1 & 2 & 3 & 4 & 5 & 6 & 7 & 8 & 9 \\
        \hline
        - & 0.22 & 0.57 & 2.85 & 2.82 & 0.28 & 6.64 & 1.41 & 2.38 & 3.14 \\
        \hline
        0.22 & - & 10.52 & 0.13 & 1.38 & 0.01 & 0.14 & 0.58 & 0.22 & 0.19 \\
        \hline
        0.57 & 10.52 & - & 1.43 & 6.97 & 0.01 & 0.18 & 19.13 & 4.48 & 0.19 \\
        \hline
        2.85 & 0.13 & 1.43 & - & 1.62 & 0.47 & 0.64 & 4.00 & 12.89 & 15.87 \\
        \hline
        2.82 & 1.38 & 6.97 & 1.62 & - & 0.31 & 2.06 & 1.42 & 1.24 & 2.52 \\
        \hline
        0.28 & 0.01 & 0.01 & 0.47 & 0.31 & - & 1.75 & 0.03 & 0.25 & 0.33 \\
        \hline
        6.64 & 0.14 & 0.18 & 0.64 & 2.06 & 1.75 & - & 0.31 & 0.87 & 0.41 \\
        \hline
        1.41 & 0.58 & 19.13 & 4.00 & 1.42 & 0.03 & 0.31 & - & 1.66 & 0.29 \\
        \hline
        2.38 & 0.22 & 4.48 & 12.89 & 1.24 & 0.25 & 0.87 & 1.66 & - & 7.81 \\
        \hline
        3.14 & 0.19 & 0.19 & 15.87 & 2.52 & 0.33 & 0.41 & 0.29 & 7.81 & - \\
        \hline
    \end{tabular}
    \label{tab:variance_accuracy_nn4eo}
\end{table*}

\begin{table*}[h]
    \centering
    \caption{HQNN4EOv3 - Variance of Test Accuracy across 10 seed}
    \begin{tabular}{|c | c | c | c | c | c | c | c | c | c|}
        \hline
        0 & 1 & 2 & 3 & 4 & 5 & 6 & 7 & 8 & 9 \\
        \hline
        - & 0.13 & 2.54 & 2.97 & 2.86 & 0.58 & 7.98 & 1.35 & 2.91 & 6.83 \\
        \hline
        0.13 & - & 4.05 & 0.14 & 1.39 & 0.00 & 0.07 & 0.68 & 0.32 & 0.21 \\
        \hline
        2.54 & 4.05 & - & 1.22 & 0.46 & 0.00 & 0.17 & 16.69 & 2.26 & 0.69 \\
        \hline
        2.97 & 0.14 & 1.22 & - & 0.95 & 0.43 & 1.45 & 1.59 & 10.05 & 7.06 \\
        \hline
        2.86 & 1.39 & 0.46 & 0.95 & - & 0.42 & 2.68 & 4.86 & 1.16 & 2.82 \\
        \hline
        0.58 & 0.00 & 0.00 & 0.43 & 0.42 & - & 2.46 & 0.03 & 0.08 & 0.10 \\
        \hline
        7.98 & 0.07 & 0.17 & 1.45 & 2.68 & 2.46 & - & 0.13 & 1.24 & 0.49 \\
        \hline
        1.35 & 0.68 & 16.69 & 1.59 & 4.86 & 0.03 & 0.13 & - & 1.54 & 0.35 \\
        \hline
        2.91 & 0.32 & 2.26 & 10.05 & 1.16 & 0.08 & 1.24 & 1.54 & - & 13.99 \\
        \hline
        6.83 & 0.21 & 0.69 & 7.06 & 2.82 & 0.10 & 0.49 & 0.35 & 13.99 & - \\
        \hline
    \end{tabular}
    \label{tab:variance_accuracy_qnn4eo}
\end{table*}

\begin{table*}[h]
    \centering
    \caption{ViT - Variance of Test Accuracy across 10 seed}
    \begin{tabular}{|c | c | c | c | c | c | c | c | c | c|}
        \hline
        0 & 1 & 2 & 3 & 4 & 5 & 6 & 7 & 8 & 9 \\
        \hline
        - & 0.49 & 10.95 & 6.44 & 1.70 & 3.24 & 3.10 & 10.08 & 2.15 & 3.27 \\
        \hline
        0.49 & - & 8.03 & 0.08 & 0.63 & 0.11 & 0.44 & 2.19 & 0.12 & 0.25 \\
        \hline
        10.95 & 8.03 & - & 0.56 & 3.21 & 0.67 & 1.08 & 5.15 & 0.58 & 1.23 \\
        \hline
        6.44 & 0.08 & 0.56 & - & 1.62 & 6.64 & 16.37 & 1.58 & 2.66 & 15.73 \\
        \hline
        1.70 & 0.63 & 3.21 & 1.62 & - & 0.67 & 0.58 & 3.64 & 0.51 & 1.01 \\
        \hline
        3.24 & 0.11 & 0.67 & 6.64 & 0.67 & - & 11.36 & 1.85 & 1.50 & 10.26 \\
        \hline
        3.10 & 0.44 & 1.08 & 16.37 & 0.58 & 11.36 & - & 0.79 & 7.04 & 2.85 \\
        \hline
        10.08 & 2.19 & 5.15 & 1.58 & 3.64 & 1.85 & 0.79 & - & 2.63 & 0.27 \\
        \hline
        2.15 & 0.12 & 0.58 & 2.66 & 0.51 & 1.50 & 7.04 & 2.63 & - & 0.97 \\
        \hline
        3.27 & 0.25 & 1.23 & 15.73 & 1.01 & 10.26 & 2.85 & 0.27 & 0.97 & - \\
        \hline
    \end{tabular}
    \label{tab:variance_accuracy_vit}
\end{table*}

\begin{table*}[h]
    \centering
    \caption{HQViT - Variance of Test Accuracy across 10 seed}
    \begin{tabular}{|c | c | c | c | c | c | c | c | c | c|}
    \hline
    0 & 1 & 2 & 3 & 4 & 5 & 6 & 7 & 8 & 9 \\
    \hline
    - & 2.29 & 7.17 & 4.87 & 1.11 & 9.53 & 1.34 & 14.31 & 2.87 & 1.51 \\
    \hline
    2.29 & - & 8.54 & 0.50 & 3.02 & 0.04 & 2.16 & 2.76 & 0.09 & 0.23 \\
    \hline
    7.17 & 8.54 & - & 1.04 & 5.25 & 2.29 & 5.66 & 3.78 & 0.71 & 0.29 \\
    \hline
    4.87 & 0.50 & 1.04 & - & 4.28 & 10.37 & 138.76 & 5.58 & 6.88 & 2.99 \\
    \hline
    1.11 & 3.02 & 5.25 & 4.28 & - & 0.83 & 1.35 & 10.02 & 0.81 & 1.02 \\
    \hline
    9.53 & 0.04 & 2.29 & 10.37 & 0.83 & - & 38.83 & 1.97 & 1.16 & 14.79 \\
    \hline
    1.34 & 2.16 & 5.66 & 138.76 & 1.35 & 38.83 & - & 1.86 & 21.12 & 3.57 \\
    \hline
    14.31 & 2.76 & 3.78 & 5.58 & 10.02 & 1.97 & 1.86 & - & 1.17 & 0.43 \\
    \hline
    2.87 & 0.09 & 0.71 & 6.88 & 0.81 & 1.16 & 21.12 & 1.17 & - & 0.70 \\
    \hline
    1.51 & 0.23 & 0.29 & 2.99 & 1.02 & 14.79 & 3.57 & 0.43 & 0.70 & - \\
    \hline
    \end{tabular}
    \label{tab:variance_accuracy_qvit}
\end{table*}

\end{document}